\documentclass[manuscript]{acmart}
\AtBeginDocument{%
  }

\setcopyright{acmlicensed}
\copyrightyear{2025}
\acmYear{2025}
\acmDOI{XXXXXXX.XXXXXXX}

\acmJournal{JACM}
\acmVolume{37}
\acmNumber{4}
\acmArticle{111}
\acmMonth{1}

\usepackage{graphicx}%
\usepackage{multirow}%
\usepackage{amsmath}

\usepackage{amssymb}
\usepackage{amsthm}%
\usepackage{mathrsfs}%
\usepackage[title]{appendix}%
\usepackage{xcolor}%
\usepackage{textcomp}%
\usepackage{manyfoot}%
\usepackage{subcaption}
\usepackage{booktabs}%
\usepackage{algorithm}%
\usepackage{algorithmicx}%
\usepackage{algpseudocode}%
\usepackage{listings}%
\usepackage{tabularx} %
\usepackage{booktabs}  
\usepackage[table,xcdraw]{xcolor}  
\usepackage{makecell}  
\usepackage{subcaption}
\usepackage{multicol} %

\begin{document}

\title{Modeling Inter-drone Interference as a Service in Skyway Networks}

\author{Gabriel Timothy}
\orcid{0009-0005-0648-5824}
\email{gabriel.timothy@sydney.edu.au}
\orcid{}
\affiliation{%
  \institution{School of Computer Science, The University of Sydney}
  \city{Sydney}
  \state{NSW}
  \postcode{2006}
  \country{Australia}
}

\author{Syeda Amna Rizvi}
\orcid{0000-0002-6844-9248}
\email{amna.rizvi@sydney.edu.au}
\affiliation{%
  \institution{School of Computer Science, The University of Sydney}
  \city{Sydney}
  \state{NSW}
  \postcode{2006}
  \country{Australia}
}

\author{Muhammad Umair}
\orcid{0000-0002-2719-4955}
\email{muhammad.umair@sydney.edu.au}
\affiliation{%
  \institution{School of Computer Science, The University of Sydney}
  \city{Sydney}
  \state{NSW}
  \postcode{2006}
  \country{Australia}
}

\author{Athman Bouguettaya}
\orcid{0000-0003-1254-8092}
\email{athman.bouguettaya@sydney.edu.au}
\affiliation{%
  \institution{School of Computer Science, The University of Sydney}
  \city{Sydney}
  \state{NSW}
  \postcode{2006}
  \country{Australia}
}

\author{Balsam Alkouz}
\orcid{0000-0001-7938-4438}
\email{balsam.alkouz@sydney.edu.au}
\affiliation{%
  \institution{School of Computer Science, The University of Sydney}
  \city{Sydney}
  \state{NSW}
  \postcode{2006}
  \country{Australia}
}
\affiliation{%
  \institution{King Abdullah University of Science and Technology}
  \city{Thuwal}
  \state{Makkah}
  \postcode{23955-6900}
  \country{Saudi Arabia}
}

\renewcommand{\shortauthors}{Timothy et al.}

\begin{abstract}
We present a novel investigation into the impact of \textit{inter-drone interference} on delivery efficiencies within \textit{multi-drone skyway networks}. We conduct controlled experiments to analyze the behavior of drones in an indoor testbed environment. Our study compares performance between solo flights and concurrent \textit{multi-drone} operations along predefined routes. This analysis captures interference occurring during both flight and at charging stations, providing a comprehensive evaluation of its effects on overall network performance. We conduct a comprehensive series of experiments across diverse scenarios to systematically understand and model the dynamics of inter-drone interference. Key metrics, such as power consumption and \textit{delivery times}, are considered. This generates a comprehensive dataset for in-depth analysis of interference at both the node and segment levels. These findings are then formalized into a predictive model. The results validate the effectiveness of the developed model, demonstrating its potential to accurately forecast inter-drone interferences.
\end{abstract}





\maketitle

\section{Introduction}\label{sec:introduction}

The rapid expansion of logistics and \textit{delivery services} has significantly strained traditional ground-based transportation methods, leading to delays and higher operational costs~\cite{2021impact}. Drones—also known as Unmanned Aerial Vehicles (UAVs)—have emerged as a promising solution for meeting the growing demand for faster and more cost-effective delivery options~\cite{akram2023chained}. Recent researches indicate that drones can reduce delivery times by up to 50\% compared to traditional ground transportation methods, while significantly reducing operational costs~\cite{McKinsey2023}. Drones facilitate efficient navigation through complex environments, making them ideal for last-mile deliveries and time-sensitive shipments, such as medical supplies \cite{betti2024uav}. This efficiency is driving a surge in demand for drone deliveries across critical sectors, consequently leading to densely populated airspace with multiple \textit{drones}. In this complex environment, \textit{inter-drone interference} can pose significant challenges to delivery \textit{efficiency} and \textit{safety}.\looseness=-1

Most discussions on drone-based delivery services emphasize point-to-point deliveries within specific geographical regions~\cite{lee2021package}. However, there are situations where drones are required to cover longer distances, necessitating frequent \textit{recharging} stops to reach their destinations~\cite{hong2018range}. These constraints make route optimization more challenging, particularly when balancing recharging stops with timely deliveries. To address this challenge, recent studies adopt a service-oriented approach by modeling Drone-as-a-Service (DaaS) within a \textit{skyway network}. A skyway network is composed of an infrastructure of interconnected \textit{nodes} and \textit{segments}. The nodes represent building rooftops that serve as strategic stopovers for drones to recharge. The segments define the airspace, incorporating predefined 3D waypoints (latitude, longitude, altitude) that guide drones as they navigate between these nodes. This service-oriented approach facilitates the path composition for long-distance drone deliveries within the network~\cite{composingdaas}.

\begin{figure*} 
    \centering
    \includegraphics[width=0.6\linewidth]{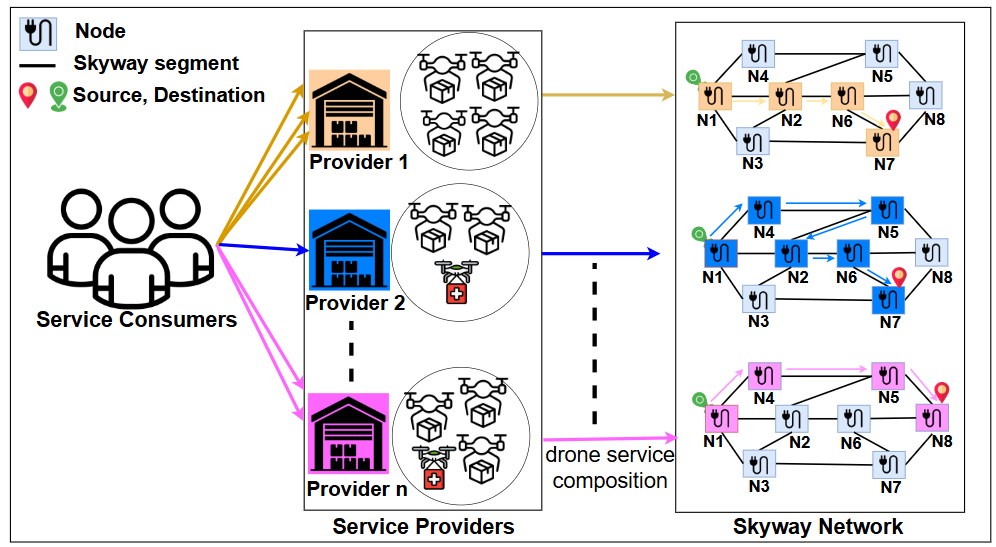}
    \caption{Path Composition by Multiple Drone Providers in a Skyway Network}
\label{fig:skywayNetworkIntro}
\end{figure*}

Operating multiple drones in close proximity within a skyway network introduces significant challenges due to interference factors~\cite{minucci2020avoiding}. As these drones navigate through crowded aerial pathways, the risk of signal disruption, collisions, and communication conflicts increases~\cite{li2021some}. Such interference can hinder the \textit{smooth} operation of flights, causing erratic movements and delays in delivery schedules~\cite{rezaee2024comprehensive}. To address these issues, we refine the service abstraction of operations within skyway networks, allowing for a more comprehensive model of inter-drone interference. This abstraction conceptualizes the skyway network as shared airspace utilized by multiple \textit{drone service providers} (i.e., drone owners). These drone owners are responsible for composing delivery paths consisting of nodes and segments to meet consumer demands for timely and cost-effective deliveries. Furthermore, a tier of a \textit{super provider} is designated to oversee the \textit{smooth}, i.e., \textit{safe} and \textit{efficient} operation of drone traffic in the shared skyway network.

Existing drone research focuses on composing drone delivery services to efficiently select and integrate the optimal service combinations for transporting goods across skyway networks~\cite{shahzaad2019constraint}. This approach involves configuring a composite route consisting of segments and intermediate nodes that the drone must traverse at defined time intervals~\cite{shahzaad2022drone}. Figure~\ref{fig:skywayNetworkIntro} illustrates an overview of path composition by multiple drone providers in a shared skyway network. Multiple service providers (i.e., drone owners) manage their own drone fleets depicted using distinct colors. Service consumers are end customers who request package delivery services from various drone providers. Each drone provider strategically plans the arrival and departure of drones from the intermediate nodes, taking into account the drones' recharging requirements to ensure seamless and timely deliveries. Drone providers compose their path \emph{independently} and \emph{greedily} with a goal of minimizing their own delivery time~\cite{attenni2023drone}~(Fig.~\ref{fig:skywayNetworkIntro}). This approach may lead to \emph{interference} as drones compete for limited recharging pads and shared airspace, potentially impacting overall delivery efficiency and safety. In this context, \textit{super providers} play a critical role, optimizing overall network utilization while assisting drone providers in maintaining their Quality of Service (QoS) requirements. 

Current solutions often overlook inter-drone interference, which can lead to congestion and delays, ultimately compromising the ability to meet QoS requirements, particularly regarding \textit{delivery times}. Most of the recent research in drone traffic management focuses on modeling interferences related to \emph{safety}, such as \emph{collisions} \cite{besada2022modelling}. However, to the best of our knowledge, no prior studies have examined how various interferences affect the \emph{efficiency} of drone services. This work is the first to comprehensively investigate the impact of inter-drone service interferences on critical performance metrics like \textit{delivery time} and \textit{energy consumption}. 
The proposed interference impact model provides a foundation for developing drone traffic management systems. These systems will support the design of interference resolution strategies, such as real-time path adjustments, to generate interference-free paths for drones. Real-time interference resolution strategies require sophisticated methods due to the shared nature of the skyway network. These methods must ensure fairness and efficiency as multiple drone services operate concurrently. For instance, a resolution strategy that benefits one drone service may negatively impact others. The development of a framework for interference-aware path adjustments is a future extension of this work.

We define \textit{interference} as an event that disrupts a drone service provider's ability to meet QoS requirements, particularly regarding \textit{delivery time}. Interferences can manifest at both the \textit{node} and \textit{segment} levels. \emph{Node-level interference} specifically refers to delays caused by disruptions at recharging stations. For instance, if recharging pads are limited, a drone may need to wait longer than expected when multiple drones are queued for access. Additionally, a drone might be forced to \textit{hover} while waiting for a pad, consuming extra battery power. In such cases, the increased energy demand or prolonged waiting times caused by interference directly compromise the drone service's ability to meet its QoS targets. \textit{Segment-level interference} refers to the impact on a drone's expected power consumption as it navigates specific segments of the skyway network. This type of interference occurs when \textit{multiple} drones operate in close spatial proximity such that their aerodynamic forces interfere. Consequently, drones may need to expend more battery power to maintain their flight paths, leading to additional recharging needs at subsequent nodes to ensure successful package delivery. For instance, a segment-level inter-drone service interference may occur between drones when multiple drones are required to operate concurrently within the same segment. In this scenario, the downwash forces generated by a drone operating at a higher altitude would increase the energy consumption of the drone flying below it~\cite{shouji2021mean}. This additional energy expenditure may require extra recharging at subsequent nodes, disrupting providers' service plans and compromising their ability to meet QoS targets for \textit{delivery time}. 

We provide a novel approach to model a drone's \textit{node-level} and \textit{segment-level interference}. This model considers key efficiency factors, including recharging pad utilization at nodes and inter-drone interactions during segment traversal. To this end, we collect and analyze a dataset to investigate the impacts of both \textit{node-level} and \textit{segment-level} interferences on the delivery times of an autonomous fleet of drone flights. Each drone flight varies according to the specific intermediate nodes it visits, depending on the designated source and destination nodes. Additionally, each drone flight differs in \textit{payload}, resulting in varying recharging requirements at the nodes. We conduct a comprehensive series of experiments involving these drone flights in an indoor testbed designed to emulate a skyway network that replicates urban conditions. We further extend the experiments to include varying \textit{wind} conditions, accounting for environmental factors that can influence both \textit{delivery time} and energy consumption. To quantify the influence of these interferences, we employ advanced empirical modeling techniques, allowing us to capture the nuanced relationships between external conditions and the efficiency of drone services. Specifically, we model the inter-drone interference using a polynomial regression model. Results validate the effectiveness of the developed empirical model. The key contributions of this work include the following:
\begin{itemize}
    \item We present a comprehensive analysis of \textit{inter-drone interferences} that occur within a \textit{multi-drone} skyway network, emphasizing the complexities introduced by concurrent operations in shared skyway networks. 
    \item We propose a model to quantify the impact of inter-drone interference, assessing its effects on operational efficiency at both the segment and node levels.
    \item We develop a Graphical User Interface to facilitate the tracking of multiple drone trajectories within a skyway network. This interface allows users to visualize real-time drone movements, enhancing operational awareness and coordination among drones. 
\end{itemize}

It is important to note that additional factors may influence inter-drone interference, such as the payload's \textit{volume} and \textit{shape}, drone's \textit{size}, and \textit{speed}. These variables, while not included in our current model, can have a significant impact on drone stability, power consumption, and overall performance. Future iterations of the model could be enhanced by incorporating these parameters, offering a deeper understanding of interference dynamics in real-world conditions. This refinement would lead to improved accuracy in predicting the performance of multi-drone systems under a broader range of environmental and operational scenarios.
\section{Motivating Scenario}
\begin{figure*}[t!]  
    \centering
    \includegraphics[width=0.9\linewidth]{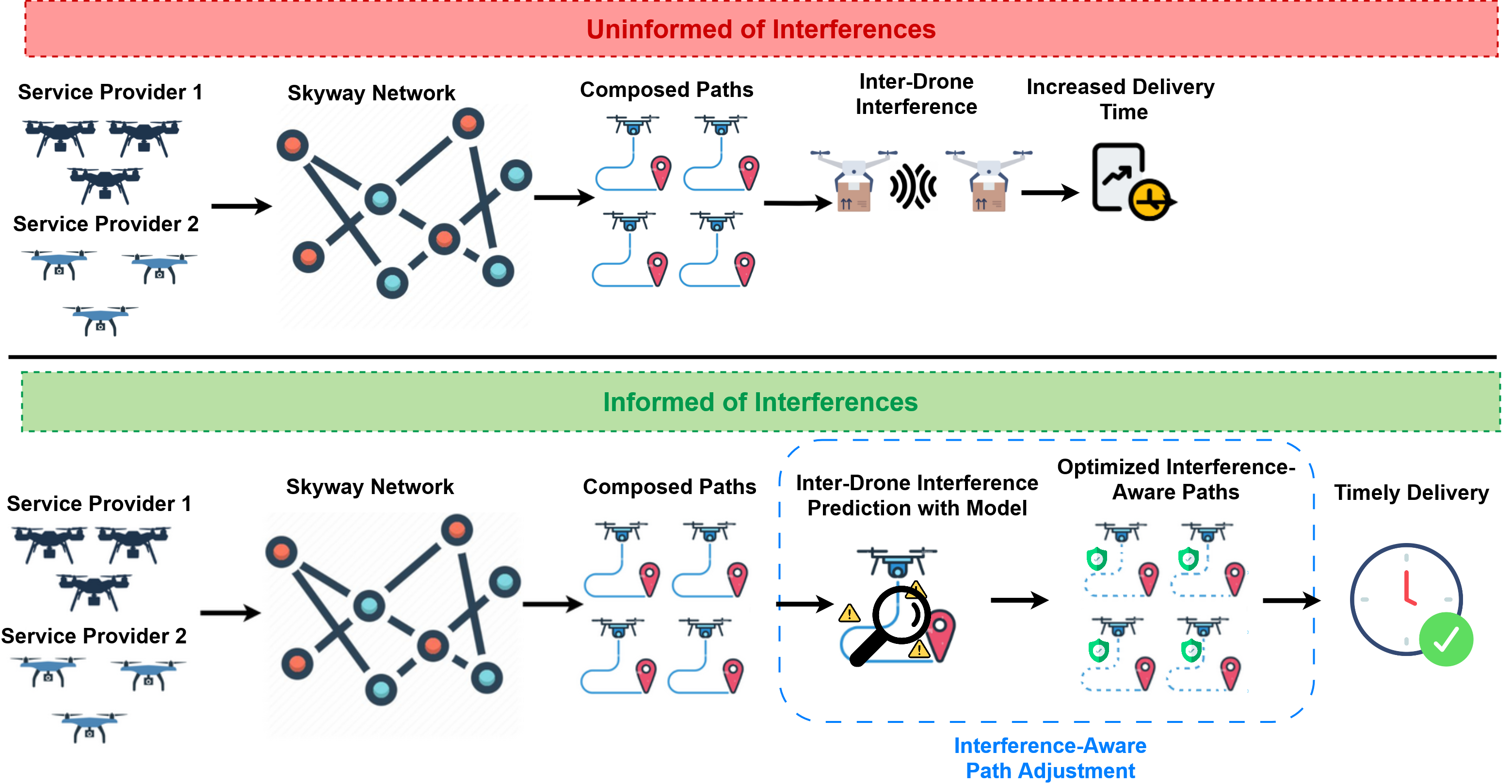}
    \caption{A Motivating Example}
    \label{fig:motivating_scenario}
\end{figure*}

As drone delivery services grow in sectors like e-commerce, healthcare, and emergency response, the challenges of coordinating multiple drones in shared airspace become more apparent. Efficient coordination is crucial to avoid operational inefficiencies, especially in densely populated areas. Inter-drone interference, defined as the disruption caused by the interaction of drones operating in close proximity, poses significant challenges to delivery \textit{efficiency} and \textit{safety}. 
To illustrate the complexities of inter-drone interference, several notable incidents underscore the challenges of operating drones in shared airspace. A prominent example is the Angle Lake drone show failure during the 2023 Fourth of July celebration in SeaTac, where 55 of 200 drones unexpectedly dropped into the lake, resulting in equipment losses estimated at \$143,000. This incident highlights vulnerabilities to external interference, potentially caused by factors such as off-frequency cell towers or unauthorized jamming devices. Another significant case occurred in Brisbane, where a delivery drone became entangled in power lines, leading to power outages for local residents. This incident reflects the navigational challenges faced by drones and the risks of inter-drone interference, particularly when multiple drones operate in close proximity. Such challenges illustrate the need for robust interference management strategies to ensure the reliability and safety of drone operations as their use expands across various sectors.

These incidents emphasize the urgent necessity of implementing efficient drone traffic management solutions to mitigate such interferences. We illustrate this motivation in Fig.~\ref{fig:motivating_scenario}, which presents a comparative example of drone path planning in a shared skyway network. As shown in Fig.~\ref{fig:motivating_scenario}, path planning without being aware of potential inter-drone interference leads to overlapping routes, resulting in delivery delays. In contrast, managing drone traffic with interference awareness allows service providers to adjust their paths proactively, thus enabling timely deliveries. Our study aims to build on the aforementioned challenges by modeling the effects of inter-drone interference, particularly focusing on aerodynamic disturbances and their impact on energy consumption and delivery times.

\section{Related Work}
We identify three key areas of research that are closely related to our work: (1) \textit{service-based} drone deliveries, (2) drone aerodynamic disturbances in proximity flight, and (3) drone recharging resource utilization and queue dynamics.

\subsection{Service-based Drone Deliveries}
The service computing framework uses an on-demand model, allowing users to access services without needing to manage the underlying resources~\cite{bouguettaya-2017}. 
Recent works leverage the service paradigm to model drones as a service in terms of functional and non-functional attributes~\cite{alkouz-2022}.
The functional aspect of drones is primarily concerned with their ability to transport packages within a defined skyway network~\cite{alkouz-2022}.
A key functional capability is the drone's ability to navigate these networks to ensure precise and timely deliveries. Non-functional capabilities, or Quality of Service (QoS) attributes, include factors such as payload capacity, battery life, flight range, and recharging needs. These attributes determine a drone's overall performance and the feasibility of specific delivery routes~\cite{alkouz-2022}. 

Due to these constraints, a single drone is often insufficient for long-range deliveries. Recent work in Drone-as-a-Service (DaaS) combines multiple drone services to meet delivery requirements beyond a single drone's capabilities~\cite{composingdaas}. Drones manage segments of the delivery route based on their capabilities. Composing drone services involves selecting the best combination of drones to meet delivery requirements, with each drone offering unique QoS attributes. The goal is to complete deliveries on time and at the lowest cost, while accounting for operational constraints like recharging points and weather conditions~\cite{composingdaas}. Recent algorithms, including heuristic and Dijkstra-based approaches, optimize this process by selecting the best drone service combination. These algorithms consider spatio-temporal factors, payload capacity, and battery life to ensure cost-effective and timely deliveries~\cite{composingdaas}. However, to the best of our knowledge, no existing framework considers the impact of inter-drone interactions, such as interference, on power consumption and delivery times. Our framework addresses this gap by providing insights into the unwarranted effects of inter-drone interference, helping service providers make more informed operational decisions.

\subsection{Aerodynamic Interference in Close-Proximity Drone Flights}
When drones operate in close proximity, they generate complex aerodynamic interactions that significantly affect flight stability and efficiency. This instability is due to the downwash effect, where airflow generated by a drone at a higher altitude impacts the drone at a lower altitude~\cite{montagner2024effects,schmidt2016stability}.
These disturbances become especially pronounced during formation flying. For example, when drones fly in a column formation, the lead drone creates a downwash that affects the performance of the trailing drones. The effect is more pronounced when drones fly at high speeds or in turbulent conditions~\cite{Guo2023}. 
The forces cause increased drag and reduced lift, leading to higher power consumption as the affected drones compensate for the disturbed airflow~\cite{zhao-2020}. Another study leverages the analysis of air disturbances around drones to detect the presence and proximity of nearby drones~\cite{zhao-2020}. The proposed method uses onboard sensors such as gyroscopes and accelerometers for low-cost proximity detection, though accuracy may vary with drone design and flight conditions~\cite{zhao-2020}. 
Recent advancements in data acquisition involve UAVs to collect information from ground-based Wireless Sensor Networks (WSNs). However, a key challenge lies in efficiently collecting this data while minimizing UAV flight time and energy consumption. A study formulates this as a UAV trajectory optimization problem to efficiently gather data from distributed ground sensors, balancing travel time, data transmission latency, and the UAVs’ energy constraints~\cite{luo-trajectory-2020}. Their work formulates and proves the underlying trajectory planning problems to be NP-Complete, and proposes approximation algorithms for real-world use. Scenario-based evaluations demonstrate the effectiveness of these algorithms in maximizing data collection efficiency while considering time and energy constraints.

Recent studies have introduced models to predict these interactions in advance. One study developed a model that estimates the forces and torques generated by two multi-rotors flying in close formation~\cite{jain-2019}. 
The model predicts aerodynamic forces, including drag, and torques that influence the drone's orientation, with outputs varying based on the drones' vertical and horizontal separation. The model is most accurate for vertical separations greater than seven times the drone's size~\cite{jain-2019}. While the proposed work effectively addresses aerodynamic interference at the micro-level (peer-to-peer drone interactions), these disturbances become significantly more complex when multiple drones operate concurrently within larger formations. Given the anticipated growth in drone traffic, managing and tracking drone formations becomes increasingly challenging. A Weighted Component Stitching (WCS) method and a weighted component-based Kalman filter algorithm (both with O(n³) complexity) are proposed to reliably track drone formations in sparse and noisy network conditions~\cite{wang-formation-tracking-2018}. Extensive experiments demonstrate that these methods improve formation tracking accuracy by 21\%–48\% compared to state-of-the-art approaches.

To the best of our knowledge, no existing model captures the cumulative effects of aerodynamic disturbances throughout an entire operational plan, rather than just isolated experiments. Our work fills this gap by examining how these disturbances impact the power consumption and delivery time of drones. This approach offers a comprehensive view of inter-drone interference, allowing for better-informed operational decisions in \textit{multi-drone} networks.

\subsection{Optimizing Drone Recharging and Queue Management}

The limited battery life of drones necessitates frequent recharging to maintain continuous operation. Typically, drone batteries support flights of approximately 25 to 30 minutes before requiring a recharge or replacement~\cite{bradley-2023, wang-2022}. As drone applications expand into domains such as e-commerce and medical deliveries, the need for robust recharging infrastructure becomes increasingly critical. 
UAVs require more frequent short-duration charging cycles that place an unprecedented strain on existing charging infrastructure~\cite{lucic-2023}. A recent study investigates the deployment of shared recharging infrastructure for UAVs and electric vehicles (EVs). Results indicate that shared usage leads to resource contention, causing delays in UAV recharging~\cite{qin-2022}. These findings imply that the current recharging infrastructure may become congested with the addition of UAV resource demand and will require enhancements to manage the increased load effectively~\cite{qin-2022}. Recent studies propose load balancing and route optimization techniques to distribute UAV traffic across stations for reducing congestion~\cite{bacanli-2022}. To the best of our knowledge, no study has explored the cumulative effects of inter-drone interference at shared charging stations. This work presents the first attempt to model how such interactions impact delivery efficiency and safety.

\section{System Design and Implementation}
\begin{figure}[t!]
    \centering
    \begin{minipage}[b]{0.49\linewidth}
        \centering
        \includegraphics[width=\linewidth]{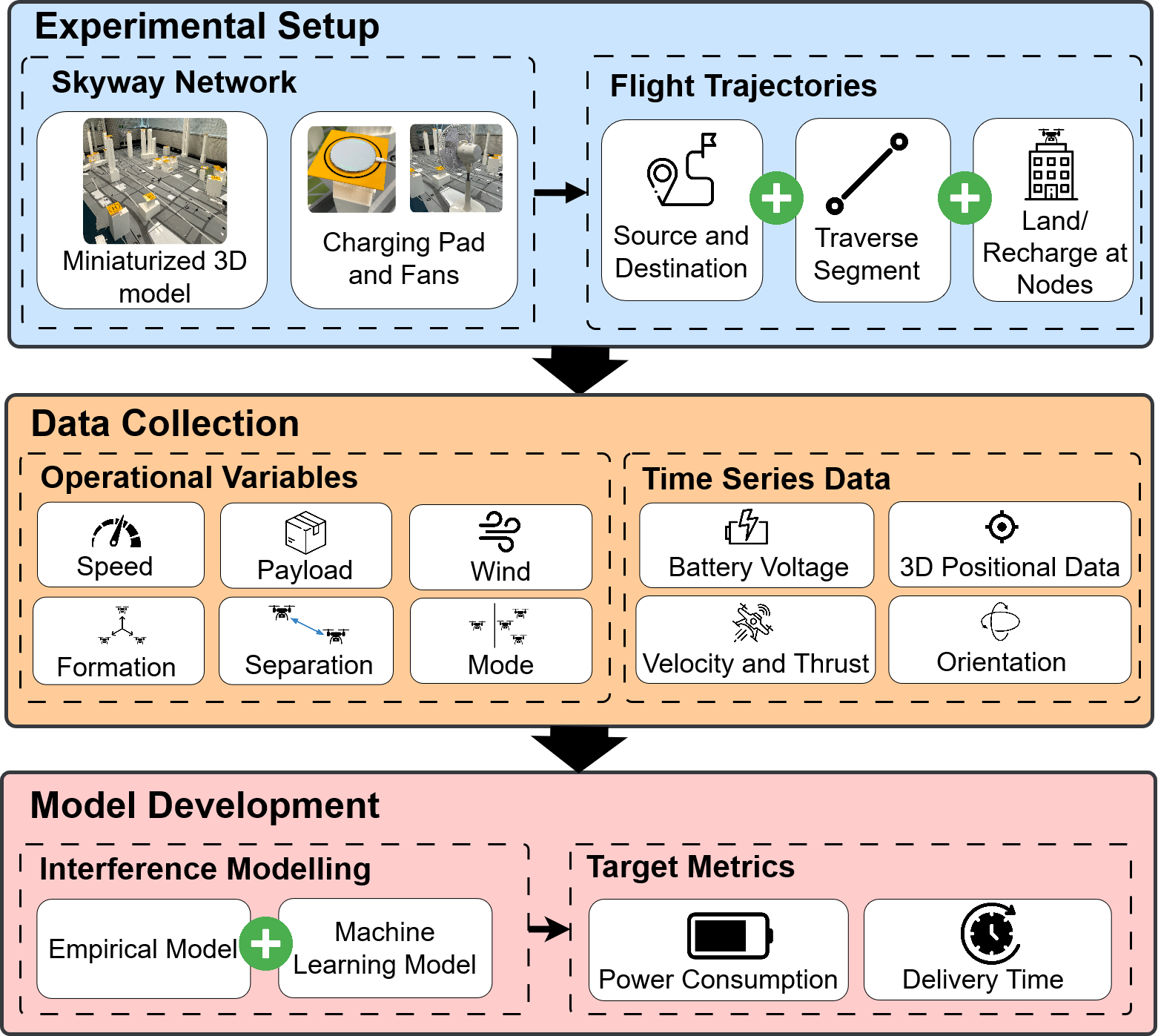}
        \caption{System Design Main Block Diagram}
        \label{fig:methodology_diagram}
    \end{minipage}
    \hfill
    \begin{minipage}[b]{0.49\linewidth}
        \centering
        \includegraphics[width=\linewidth]{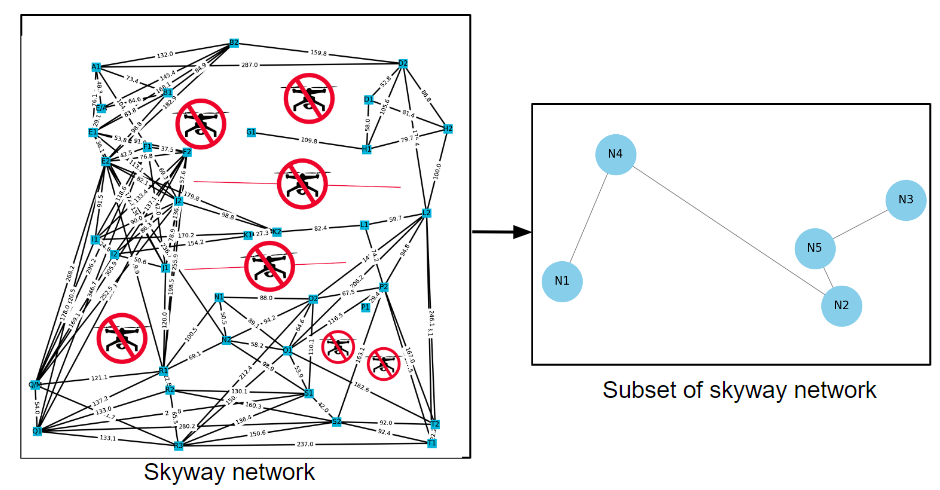}
        \caption{Skyway Network~\cite{lee-2021}}
        \label{skyway network}
    \end{minipage}
\end{figure}

We built our system to conduct an in-depth analysis of inter-drone interference within a multi-drone skyway network. Figure~\ref{fig:methodology_diagram} illustrates the overall workflow of our approach, including the experimental setup, data collection, and model development stages. We performed a comprehensive data collection of autonomous drone delivery flights in an indoor drone testbed. This indoor drone testbed consists of a skyway network based on the 3D model of Sydney CBD~\cite{Guo2023}. We consider a subset of the skyway network consisting of nodes (i.e., building rooftops) interconnected with Line of Sight (LoS) segments as shown in Fig.~\ref{skyway network}. We collected time-series data for each drone flight when it operates individually and simultaneously with peer drones in the data collection phase. In the development stage, we use the collected data to develop empirical and machine learning models for the interferences occurring at the node and segment level, focusing on power consumption and delivery time as key performance metrics. Finally, we evaluate the model's performance to ensure its applicability in larger networks and under varying conditions.

\subsection{Data Collection Protocol}
To ensure the reliability and accuracy of our model, we adopted a controlled data collection protocol for producing a high quality dataset suitable for modeling. We exclude confounding factors that are inherent to drones, which may influence the impact of proximity-induced interferences on drones' energy consumption. These factors include drones' speed, payload volume, and drone model~\cite{beigi2022overview}. For instance, varying drone speed and drone proximity may vary the energy consumption of the drone. This makes it challenging to isolate whether the observed energy variation is due to inter-drone interference or the change in speed, and to what extent each factor contributes. Therefore, we fixed the speed across all drone flights, with uniform payloads, and Crazyflie 2.1 as the drone model.

We setup an indoor drone testbed, as conducting outdoor experiments involving simultaneous flights of commercial drone models is restricted under government regulations~\cite{report}. We use Crazyflie 2.1 drone model due to its suitability for indoor experiments. These drones share the same multi-rotor architecture of different commercial drone models (e.g., DJI Flycart 30, Amazon MK30) used by companies such as Amazon and DHL~\cite{airservices}. Therefore, they exhibit similar aerodynamic behavior due to the continuous rotation of propellers. However, the severity of interference impact would scale with propeller size, as larger drones produce stronger turbulent forces~\cite{hage2023investigating}. We emulate dynamic drone flights considering various parameters that closely mimic practical scenarios. These parameters include varying drone positions, formations, wind speed, wind direction, and payloads by weight. These factors allow us to systematically capture the effects of inter-drone interference on drone delivery efficiency and safety. Note that factors such as payload volume, drone speed, and model are outside the scope of our current model. Extending our model to heterogeneous fleets of drones, considering these factors, remains an important direction for future work.

\subsection{Data Collection Strategy}\label{datacollectionstrategy}
We collected a real dataset of multiple drone flights to analyze the impact of node and segment-level interferences on drone delivery efficiency. The dataset includes two types of data: \emph{solo drone flights}, where drones operate independently, and \emph{simultaneous drone flights}, where multiple drones operate simultaneously within a multi-drone skyway network. This dataset records the impact of delays provoked by the peer drones in a multi-drone skyway network in terms of delivery time and energy consumption. 
The collected dataset contains various parameters of a drone when it operates \emph{solo} and when it operates \emph{simultaneously} with peer drones. The details of these parameters are listed in Table~\ref{tab:data_parameters}.

\begin{table}[htbp]
    \small
    \rowcolors{1}{white}{gray!15}  
    \caption{Parameters Collected During Data Collection}
    \begin{tabularx}{0.9\textwidth}{l X} 
        \toprule
        \textbf{Parameter}  & \textbf{Description and Varied Values} \\ 
        \midrule
        Battery Voltage Level     & Voltage of the drone's battery (V)  \\ 
        3D Positional Data        & x, y, z coordinates of the drone, sampled every 0.1s \\ 
        Orientation Data          & Roll, Pitch, Yaw angles (degrees) \\ 
        Wind                      & Direction and intensity of wind (None, Light/Intense Headwind, Light/Intense Tailwind) \\ 
        Payload                   & Weight carried by the drone (ranging from 0 to 4.5 grams with an increment of 1.5)
        \\ 
        Drone Formation           & Spatial arrangement of drones (Front-back, Side-by-side, Top-down) \\ 
        Drone Separation          & Inter-drone distance varied from 30~cm to 150~cm with 10~cm increments.
        \\ 
        Flight Time               & Total time to complete the flight (seconds) \\ 
        Recharging Time           & Time spent recharging at nodes (seconds) \\ 
        Delivery Time             & Total time from departure to delivery (seconds) \\ 
        \bottomrule
    \end{tabularx}
    \label{tab:data_parameters}
\end{table}

\subsubsection{Solo Drone Flights}
\begin{figure}[t!]
    \centering
    \includegraphics[width=0.6\columnwidth]{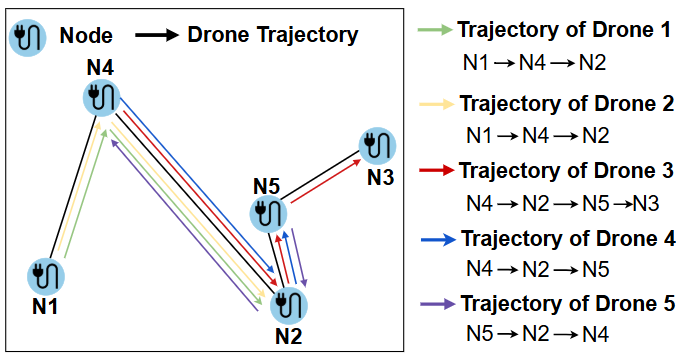}
    \caption{Trajectories of Autonomous Drone Delivery Flights}
    \label{fig:flight_plan}
\end{figure}

We autonomously directed a Crazyflie drone to carry a varying set of payload weights from the source to the destination node on a predefined trajectory. This autonomous solo flight data serves as a baseline to measure standard delivery performance without the impact of interference from peer drones. Each drone trajectory is defined based on the shortest path from the source node to the destination node (Fig.~\ref{fig:flight_plan}). However, the drone typically recharges at intermediate nodes to reach the destination node successfully due to the limited flight time of seven minutes. 
We compute the charging time of each drone at the subsequent node as follows:
\begin{equation}\label{Node_time}
    t_{node} = -\tau \cdot \ln\left(\frac{1 - S_f}{1 - S_i}\right) 
    = - (T_{full} \cdot \beta) \cdot \ln\left(\frac{1 - S_f}{1 - S_i}\right)
\end{equation}
where $S_i$ is the initial state of the drone's battery power once it reaches the subsequent node and  $S_f$ is the final state of the drone's battery power $(i.e., \approx100\%)$ after it takes off from the intermediate node. $\tau$ is the function defined in terms of $T_{full}$ and $\beta$,
where $T_{full}$ is the total time required for the drone to fully recharge, typically measured from 0\% to 99\% battery capacity, and $\beta$ is a scaling factor that accounts for variations in the charging rate as the battery charging typically follows a non-linear curve. The initial state of the drone's battery power $(S_i)$ once it reaches the subsequent node varies for each drone flight based on the payload and the distance it traverses in a segment. We repeated this set of autonomous drone flights under varying wind speeds (ranging from 6.1 km/h to 7.6 km/h with an increment of 1.5 km/h) and directions (i.e., headwinds and tailwinds). We used the time series data collected at an interval of 0.1 seconds for each drone flight to compute the total power consumption (\%) and delivery time duration from source to destination node. We compute the total power consumption $P_{total}$ for each drone flight based on the cumulative power consumed as it traverses each $i^{th}$ segment in time~$t_{\text{segment}_i}$ using Eq.~(\ref{total_Power}). 
\begin{equation}\label{total_Power}
    P_\text{{total}}=\sum_{i=1}^{n} P_i (t_{\text{segment}_i})
\end{equation}
The total power consumption~$P_{total}$ is the sum of the power consumed across all
$n$ segments of the drone's trajectory from the source node to the destination node. The total flight duration is the cumulative sum of segment traversal times across $n$ number of segments and the time spent by the drone at $m$ number of nodes to serve the delivery request. We compute total delivery duration $(DT_{total})$ using Eq.~(\ref{Delivery_Duration}) as follows:
\begin{equation}\label{Delivery_Duration}
    DT_{\text{total}} = \sum_{i=1}^{n} t_{\text{segment}_i} + \sum_{j=1}^{m} t_{\text{node}_j}
\end{equation}

\subsubsection{Simultaneous Drone Flights}
In this set of experiments, we collected the dataset of drones operating \emph{simultaneously} in the shared skyway network. We directed the same set of drone flights from their respective source to the destination node (Fig.~ \ref{fig:flight_plan}). We evaluate the impact of segment-level interferences for varying inter-drone separations of 30 cm to 150 cm with an increment of 10 cm. We allocate the recharging pad, waiting pad, and hovering zone to a drone arriving at the node based on a First-Come-First-Served (FCFS) approach. Let's assume a node $N$ with $n$ number of recharging and $m$ number of waiting pads, then the allocation process A of drone $D_i$ is defined as follows. 
\small
\begin{equation}
\begin{aligned}
A(D_i) = 
\begin{cases} 
\text{Recharging Pad}, & \text{if } i \leq n, \\
\text{Waiting Pad}, & \text{if } n < i \leq (n + m), \\
\text{Hovering Zone}, & \text{if } (n + m) < i \leq (n + m + n_{hover}),
\end{cases}
\end{aligned}
\end{equation}
\normalsize
We use the collected dataset to compute the delivery time duration using Eq.~(\ref{Delivery_Duration}) from the source to the destination node. The total power consumption is based on the power consumed while hovering at the node and during the segment. Assume a drone visits $n$ number of segments and $m$ number of nodes where the $P_j (t_{\text{hover}_j})$ is the power consumed while hovering at node $j$ for time  $t_{\text{hover}_j}$ then the total power consumption for the drone is computed as follows: 
\begin{equation}
   P_{\text{total}} = \sum_{i=1}^{n} P_i (t_{\text{segment}_i}) + \sum_{j=1}^{m} P_j (t_{\text{hover}_j})
\end{equation}

\subsection{Interference Detection and Modeling}
The detection and modeling of \emph{node} and \emph{segment-level interferences} are critical for analyzing the impact on drone delivery efficiency. Our framework processes time-series data collected from \emph{solo} and \emph{simultaneous} drone flights to \emph{detect} and \emph{quantify} the impact of inter-drones interferences on drones' delivery time. We use this data to construct mathematical models that allow us to estimate additional power consumption and delays given a drone delivery flight in a multi-drone skyway network.

\subsubsection{Modeling Segment-level Interference Effect on Power Consumption}\label{SegModel}
\begin{figure}[t!]
    \centering
    \begin{minipage}[b]{0.48\columnwidth}
        \centering
        \includegraphics[width=\linewidth]{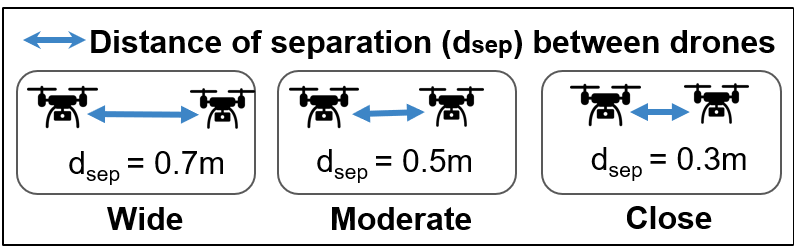}
        \caption{Drone Separation Distances used in the Study}
        \label{fig:drone_distances}
    \end{minipage}
    \hfill
    \begin{minipage}[b]{0.48\columnwidth}
        \centering
        \includegraphics[width=\linewidth]{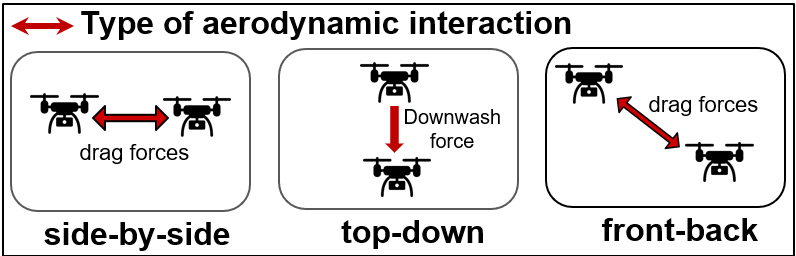}
        \caption{Different Drone Formations used in the study}
\label{fig:formation_interactions}
    \end{minipage}
    
\end{figure}

We model the impact of segment-level inter-drone interferences on the power consumption of drones. This impact informs the effect on the delivery time of drones. We identify several factors that would impact the expected power consumption of drones in a segment. These factors include \textit{payload weight}, \textit{drone separation}, \textit{drone formation}, \textit{wind conditions}, and \textit{drone position}. We formally define the aforementioned factors as follows:
\\
\textbf{Drone Payload:} The payload factor represents the weight carried by each drone. The payload directly impacts the power consumption of a drone. Heavier payloads increase the power needed to counteract gravitational forces and stabilize against potential aerodynamic interferences. We classify payload weights into three discrete categories of \textit{None}, \textit{Light}, and \textit{Heavy} based on the weight capacity of our drone as follows:
\begin{equation}
P_{\text{payload}} =
  \begin{cases} 
   \text{None} & \text{if payload} = 0 \, \text{g} \\
   \text{Light} & \text{if payload} = 3 \, \text{g} \\
   \text{Heavy} & \text{if payload} = 4.5 \, \text{g}
  \end{cases}
\end{equation}
\\
\textbf{Drone Separation:} This factor refers to the spatial separation between drones in a 3D segment.  Let \( (x_1, y_1, z_1) \) and \( (x_2, y_2, z_2) \) be the coordinates of two drones at time \( t \) then the separation distance $d_{sept}(t)$ at any time $t$ is computed as:
\small
\[
d_{\text{sep}}(t) = \sqrt{(x_1(t) - x_2(t))^2 + (y_1(t) - y_2(t))^2 + (z_1(t) - z_2(t))^2}
\]
\normalsize
This factor plays a crucial role in modeling segment-level interferences, as the close proximity between drones may indicate a significant impact on the battery power consumption of drones. We collected the dataset for varying distances of separation between drones, ranging from 30cm to 150cm with an increment of 10cm, as illustrated in Fig.~\ref{fig:drone_distances}. We categorize these distances of separation as wide, moderate, and close as follows: 
\begin{equation}
\text{Category}(d_{\text{sep}}) =
  \begin{cases} 
   \text{Close} & \text{if } 0.3 \, \text{m} \leq d_{\text{sep}} < 0.5 \, \text{m} \\
   \text{Moderate} & \text{if } 0.5 \, \text{m} \leq d_{\text{sep}} < 0.7 \, \text{m} \\
   \text{Wide} & \text{if } d_{\text{sep}} \geq 0.7 \, \text{m}
  \end{cases}
\end{equation}
\\
\textbf{Drone Formation:}
This factor refers to the spatial arrangement of the drone relative to the peer drone. We determine the formation of the drone relative to the peer drone based on the distance between the drones along the 3D axis. We determine the formation of a drone relative to its peer drone as,
\begin{equation}
f_{formation} =
  \begin{cases} 
   \text{Side-by-Side} & \text{if } |z_1 - z_2| < \delta_z \text{ and } |y_1 - y_2| < \delta_y \text{ and } |x_1 - x_2| > \delta_x \\
   \text{Top-Down} & \text{if } |x_1 - x_2| < \delta_x \text{ and } |y_1 - y_2| < \delta_y \text{ and } |z_1 - z_2| > \delta_z \\
   \text{Front-Back} & \text{if } |z_1 - z_2| < \delta_z \text{ and } |y_1 - y_2| < \delta_y \text{ and } |x_1 - x_2| > \delta_x \\
   \text{None} & \text{otherwise}
  \end{cases}
\end{equation}
where, $\delta_x$, $\delta_y$ and $\delta_z$ are the thresholds to determine the distances. Moreover, the drone formation relative to the peer drone determines the type of aerodynamic interaction between drones (Fig.~\ref{fig:formation_interactions}). For instance, in a top-down formation, the downwash forces generated from the higher-altitude drone may interfere with the drone operating in the region below it. Similarly, in a front-back formation, the trailing drone may experience turbulence due to the trailing vortices from the propellers of the leading drone \cite{Srithar2024}.  
\\
\textbf{Drone Position:} Another factor is the position of the drone relative to the peer drone, given the spatial arrangement of drones in the 3D segment. The position of the drone relative to the peer drone informs the peer-to-peer impact on the battery consumption of drones due to interference. We define the relative position of the drone given the formation of the drone in a 3D segment as:
\begin{equation}
    P_{\text{pos}} =
  \begin{cases} 
   \text{Top} & \text{if } f_{\text{formation}} = \text{Top-Down} \text{ and } z_1 > z_2 \\
   \text{Down} & \text{if } f_{\text{formation}} = \text{Top-Down} \text{ and } z_1 < z_2 \\
   \text{Left} & \text{if } f_{\text{formation}} = \text{Side-by-Side} \text{ and } x_1 < x_2 \\
   \text{Right} & \text{if } f_{\text{formation}} = \text{Side-by-Side} \text{ and } x_1 > x_2 \\
   \text{Front} & \text{if } f_{\text{formation}} = \text{Front-Back} \text{ and } x_1 > x_2 \\
   \text{Back} & \text{if } f_{\text{formation}} = \text{Front-Back} \text{ and } x_1 < x_2 \\
  \end{cases}
\end{equation}
\\
\textbf{Wind Conditions:}
We consider another factor, wind condition and spee,d that may impact the magnitude of aerodynamic interferences between drones operating in the shared segment. We collected the dataset for varying wind speeds (ranging from  6.1 km/h to 7.6 km/h with an increment of 1.5 km/h) and wind directions (i.e., headwinds and tailwinds). Wind conditions \( C_{\text{wind}} \) are defined as:

\begin{equation}
C_{\text{wind}} = \begin{cases} 
C_{\text{none}} & \text{if no wind} \\
C_{\text{light-headwind}} & \text{if light headwind} \\
C_{\text{light-tailwind}} & \text{if light tailwind} \\
C_{\text{intense-headwind}} & \text{if intense headwind} \\
C_{\text{intense-tailwind}} & \text{if intense tailwind}
\end{cases}
\end{equation}
We quantify the \emph{total segment-level interference effect} ($\hat{I}_{\text{segment}}$) based on these factors as:
\[
\hat{I}_{\text{segment}} = \beta_1 \cdot W_{\text{payload}} + \beta_2 \cdot f_{\text{formation}} + \beta_3 \cdot C_{\text{wind}} + \beta_4 \cdot d_{\text{sep}} + \beta_5 \cdot P_{\text{pos}}
\]
where the coefficients  $\beta_1$, $\beta_2$, $\beta_3$, $\beta_4$, and $\beta_5$  represent the amount of influence each factor has on the overall interference.
This interference effect represents the increase in power consumption due to the impact of peer drones operating in the same segment. We define the impact on the power consumption of drones due to inter-drone interferences and varying wind conditions in (\ref{powerconsump}).
\begin{equation}\label{powerconsump}
   P_{\text{interference}} = P_{\text{expected}} \times \hat{I}_{\text{segment}} 
\end{equation}
where \( P_{\text{expected}} \) is the power consumption without interference (i.e., baseline power consumption) and \( I_{\text{total}}\) represents the ratio of increased power consumption due to interference.
\subsubsection{Modeling Node-level Interference Effect on Delays and Power Consumption}\label{empirical model}
\label{sec:node_level_interference}

We model the impact of node-level interference on the delivery time of the drone based on the additional wait time and power consumption due to the unavailability of the recharging pad. This additional wait time $W_d$ is based on the time spent at the waiting pad $W_p$, time spent in the hovering zone $H_d$, and the additional time $T_{hover}$ to regain the power consumed while hovering. 

\begin{equation}
    W_d = H_d + W_{p,d}+ T_{hover}  
\end{equation}
where $H_d$ is the time spent in the hovering zone when no waiting pads are available and $W_{p,d}$ is the time waiting on a waiting pad before accessing a recharging pad. $T_{hover,d}$ is the additional recharging time required by drone $d$ to regain the battery power consumed while hovering for duration $H_d$. Let's assume a node with $m$ total number of recharging pads and $n$ number of total waiting, and $p$ is the number of drones ahead of drone $d$ on the waiting pads then the hovering time and waiting pad time is computed using Eq.~(\ref{hoveringT}) and Eq.~(\ref{waitingT}), respectively.
\begin{equation}\label{hoveringT}
H_d = \max \left( 0, \sum_{i=1}^{n} T_{w,i} - \sum_{j=1}^{m} T_{f,j} \right)
\end{equation}
where $T_{w,i}$ is the waiting time of each drone currently on a waiting pad to occupy the recharging pad. $T_{f,j}$ is the remaining time of each drone on the recharging pads. The waiting pad time is computed using the following formula:
\begin{equation}\label{waitingT}
W_{p,d} = \max \left( 0, \sum_{k=1}^{p} T_{r,k} - \sum_{l=1}^{m} T_{f,l} \right)    
\end{equation}
where $T_{r,k}$ is the recharging time of the $k^{th}$ drone ahead of drone $d$ in the queue for recharging and $T_{f,l}$ is the time remaining for the drones currently using the recharging pads. The additional recharging time due to hovering required by the drone $d$ with a recharging rate of $P_{recharge}$  is computed as follows:
\begin{equation}
    T_{hover}=\frac{P_{hover}*H_d}{P_{recharge}}
\end{equation}
\subsubsection{Modeling Interference Effect on Delivery Time}
We model the impact of both segment and node-level interferences on the estimated delivery time of drones. The impact on delivery time is based on the cumulative effect of interferences occurring across nodes and segments. Interferences at nodes typically occur due to the unavailability of recharging or waiting pads. This unavailability causes the drone to either wait or consume extra battery power while hovering. In contrast, segment-level interference impacts the drone's power consumption during traversal between nodes due to the presence of peer drones. Therefore, we formulate the impact on delivery time in Eq.~(\ref{DT_Interference}) based on two factors: (1) the cumulative \emph{wait time} due to node-level interference occurring at $n$ nodes visited by the drone (2) The additional time to recharge due to segment-level interference while traversing m segments. We formulate the impact on delivery time based on the cumulative effects of interferences across the entire flight path as follows:

\begin{equation}\label{DT_Interference}
      DT_{interference} = DT_{expected} + \sum_{i=1}^{n} (W_d) + \sum_{j=1}^{m} T(P_{\text{interference},j})
\end{equation}
where $DT_{expected}$ is the delivery time of the drone without the impact of interferences. 

\begin{figure}[htbp]
    \centering
    \includegraphics[width=\columnwidth]{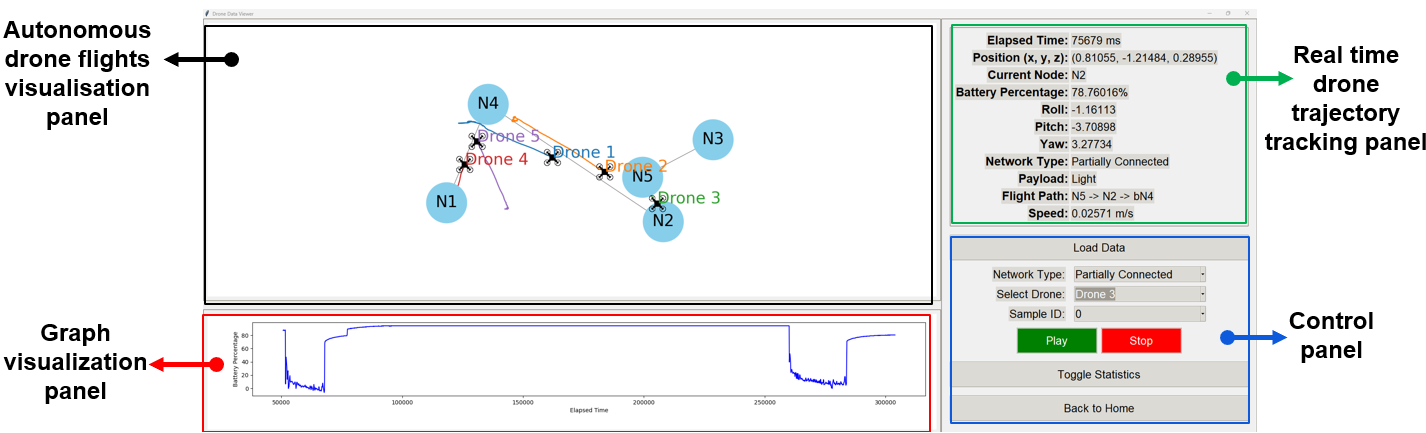}
    \caption{GUI Interface}
    \label{fig:GUI}
\end{figure}

\subsection{Graphical User Interface}
To facilitate the analysis of experimental data, we develop a Graphical User Interface (GUI) using Tkinter\footnote{https://docs.python.org/3/library/tkinter.html} (Fig. \ref{fig:GUI}). This GUI enables retrospective viewing of drone trajectories by replaying flights and reviewing key metrics over time using recorded time series data. After each experimental flight, time series data including positional coordinates (x, y, z), speed, and other flight parameters can be visualized within the interface. The GUI offers a playback feature that allows users to observe drone movements in the skyway network at different points in time. We analyze the impact of various factors, as discussed in Section~\ref{SegModel}, on the delivery time and power consumption of drones operating in a shared skyway network. \looseness=-1

\subsection{Model Development}
In this section, we aim to develop segment and node-level interference models for accurate estimation of the impact on delivery time and power consumption under various conditions. The models target two primary sources of interference: aerodynamic disturbances in flight \emph{segments} and resource contention at \emph{nodes}. For \emph{segment-level interference}, machine learning is developed to model non-linear relationships between feature variables such as drone separation, payload, and wind conditions. For \emph{node-level interference}, an empirical model is proposed in Section~\ref{empirical model} to estimate waiting and hovering times due to congestion at recharging pads. In what follows, we describe the development of a segment-level interference model.
\subsubsection{Segment-Level Interference Model}

\begin{figure}[htbp]
    \centering
    \includegraphics[width=0.5\columnwidth]{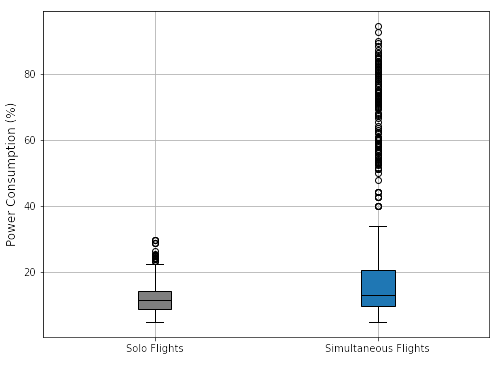}
    \caption{Boxplot comparing power consumption in solo versus simultaneous flights.}
    \label{fig:solo_vs_simultaneous_boxplot}
\end{figure}

The dataset for the segment-level interference model was generated from controlled drone flight experiments, with power consumption recorded during both solo and simultaneous flights. These flights were directed under varying conditions outlined in Section~\ref{datacollectionstrategy}. The primary goal is to quantify the impact of aerodynamic disturbances on power consumption during simultaneous flights, compared to solo flights. Figure~\ref{fig:solo_vs_simultaneous_boxplot} shows the distribution of power consumption values in solo and simultaneous flights. We observe that simultaneous flights generally exhibit higher power consumption. It is due to aerodynamic interferences between peer drones, which cause the drones to consume additional battery power for maintaining flight stability. 

\begin{figure}[htbp]
    \centering\includegraphics[width=1\columnwidth]{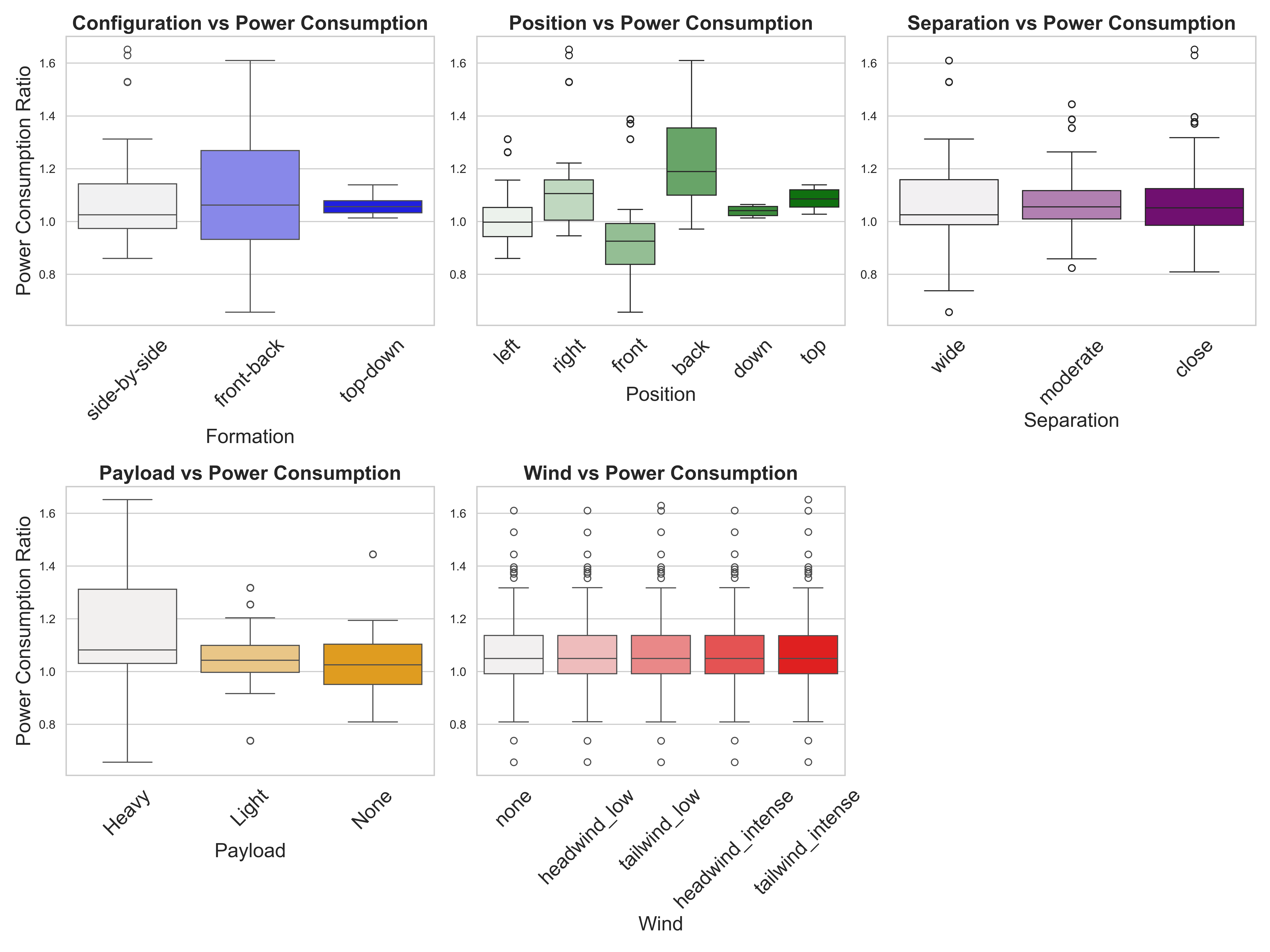}
    \caption{Boxplot showing the distribution of power consumption for different variables.}
    \label{fig:boxplot_variables}
\end{figure}

We use  \emph{interference ratio (IR)} defined in Eq.~(\ref{interferenceratio}) as a target variable in the segment-level interference model. This ratio quantifies the proportional increase in power consumption during simultaneous flights $P_{simultaneous}$ relative to solo flights $P_{solo}$. This reflects the additional energy demand caused by aerodynamic disturbances generated by the peer drones. The interference ratio forms the basis for building a predictive model using machine learning techniques.
\begin{equation}
IR = \frac{P_{\text{simultaneous}}}{P_{\text{solo}}}
\label{interferenceratio}
\end{equation}
In the next step, we analyze the relationship between feature variables (i.e., configuration, position, separation, payload, and wind) and the target variable (i.e., interference ratio). Fig. \ref{fig:boxplot_variables} illustrates that the power consumption increases for varying distances of separation between drones, payloads, and drone configurations. However, some of the feature variables exhibit a non-linear relationship with the target variable.
Therefore, we select a \emph{polynomial regression model} for accurate modeling of linear and non-linear relationships between feature variables and the target variable. We then systematically tested the polynomial degrees ranging from 1 to 4 to ensure the
appropriate polynomial degree for each variable. We use three evaluation metrics: Cross-validated Mean Absolute Error (MAE), Akaike Information Criterion (AIC), and Bayesian Information Criterion
(BIC) to assess models' performance for varying polynomial degrees. Previous studies have demonstrated the efficacy of using the aforementioned metrics in selecting the best model while avoiding the overfitting problem~\cite{wang2023genetic}. This is because the AIC and BIC metrics consider both the model complexity and fitting accuracy, with lower values indicating a better fitting model. Similarly, MAE indicates how closely the model approximates true values based on the average prediction error.

\begin{algorithm}[htbp]
\caption{Polynomial Degree Selection for Segment-Level Interference Model}
\label{alg:degree_selection}
\begin{algorithmic}[1]
\State \textbf{Input:} Dataset $\mathcal{D} = \{X, y\}$, Variables $V = \{v_1, v_2, \dots, v_n\}$
\State \textbf{Output:} Optimal degrees $\{d_{v_1}, d_{v_2}, \dots, d_{v_n}\}$ for each variable
\State Initialize polynomial degree set $D = \{1, 2, 3, 4\}$
\State Set initial degrees $d_{v} = 1$ for all $v \in V$
\For{each $v \in V$}
    \For{each $d \in D$}
        \State Apply polynomial transformation of degree $d$ to $v$
        \State Keep all other $v \in V \setminus v$ at degree 1
        \State Train model on $\mathcal{D}$
        \State \textbf{Evaluate:} Compute MAE, AIC, and BIC
        \State Normalize MAE, AIC, BIC: $\hat{MAE}, \hat{AIC}, \hat{BIC}$
        \State Compute combined score: $S(d) = \frac{\hat{MAE} + \hat{AIC} + \hat{BIC}}{3}$
        \If{$S(d) <$ Best Combined Score}
            \State Update $d_v = d$
        \EndIf
    \EndFor
\EndFor
\State Train final model with $\{d_{v_1}, d_{v_2}, \dots, d_{v_n}\}$
\State \textbf{Return:} $\{d_{v_1}, d_{v_2}, d_{v_3}, \dots, d_{v_n}\}$
\end{algorithmic}
\end{algorithm}

Algorithm \ref{alg:degree_selection} provides the details of our polynomial degree selection process. For each variable, we systematically tested polynomial degrees ranging from 1 to 4, while keeping all other variables at degree 1 (Lines 1-8). The model was trained on the dataset, and for each degree, we computed the evaluation metrics (i.e., MAE, BIC, and AIC) (Lines 9-11).  We then normalize each metric before computing a combined score for each degree due to differences in scale (Lines 11-12).
We select the polynomial degree that minimizes the combined score for each variable (Lines 13-14). Our selection results are presented in Table~\ref{table:best_degrees}.

\begin{table}[t]
    \centering
    \rowcolors{1}{white}{gray!15}
    \caption{Best polynomial degrees for each variable based on MAE, AIC, BIC, and normalized mean score.}
    \label{table:best_degrees}
    \begin{tabular}{%
        >{\raggedright\arraybackslash}p{1.5cm}   
        >{\centering\arraybackslash}p{1.8cm}     
        >{\centering\arraybackslash}p{1.5cm}     
        >{\centering\arraybackslash}p{1.5cm}       
        >{\centering\arraybackslash}p{1.5cm}       
        >{\centering\arraybackslash}p{4cm}       
    }
        \toprule
        \textbf{Variable}  & \textbf{Best Degree} & \textbf{MAE} & \textbf{AIC} & \textbf{BIC} & \textbf{Normalized Mean Score} \\ 
        \midrule
        Formation  & 1 & 0.096 & -700.857 & -640.775 & 0.143 \\ 
        Payload    & 1 & 0.092 & -700.857 & -640.775 & 0.001 \\ 
        Position   & 3 & 0.096 & -700.857 & -640.775 & 0.127 \\ 
        Separation & 4 & 0.096 & -700.857 & -640.775 & 0.126 \\ 
        Wind       & 1 & 0.093 & -700.857 & -640.775 & 0.038 \\ 
        \bottomrule
    \end{tabular}
\end{table}
We then train the polynomial regression model on the collected dataset to derive a segment-level interference model for estimating the impact of interferences on power consumption. The segment-level interference impact expression for a 4th-order polynomial is given by:
\small
\label{eq:regression_model}
\begin{align*}
\hat{I}_{\text{segment}} = &\ \beta_1 \cdot P_{\text{pos}}^3 
+ \beta_2 \cdot d_{\text{sep}}^4 
+ \beta_3 \cdot f_{\text{formation}} 
+ \beta_4 \cdot C_{\text{wind}} 
+ \beta_5 \cdot W_{\text{payload}} 
\end{align*}
\normalsize
where $\beta_1$, $\beta_2$,$\beta_3$, $\beta_4$, and $\beta_5$ are coefficients that indicate the contribution of each variable on the impact of interference. We determine the values of these coefficients based on the relative importance of each variable in predicting the interference. The relative importance of each variable was determined by summing the absolute values of its polynomial terms across all degrees. These sums were then expressed as percentages of the total contribution from all variables, yielding a measure of relative importance. As shown in Table~\ref{table:relative_impact}, the relative positions of drones to one another and their spatial proximity are key factors that influence the impact of interference. The resulting polynomial regression model for segment-level interference is expressed as follows:
\small
\label{eq:regression_model}
\begin{align*}
\hat{I}_{\text{segment}} = &\ 0.4085 \cdot P_{\text{pos}}^3 
+ 0.3272 \cdot d_{\text{sep}}^4 
+ 0.1262 \cdot f_{\text{formation}} 
+ 0.0780 \cdot C_{\text{wind}} 
+ 0.0601 \cdot W_{\text{payload}} 
\end{align*}
\normalsize
where \( \hat{I}_{\text{segment}} \) represents the predicted interference ratio, and the higher order terms for position and separation capture the non-linear effects observed in the data. 

\begin{table}[htbp]
    \centering
    \caption{Relative Impact of Variables on Interference Ratio.}
    \label{table:relative_impact}
    \begin{tabular}{lccccc}
        \toprule
        \textbf{Metric} & \textbf{Position} & \textbf{Separation} & \textbf{Formation} & \textbf{Wind} & \textbf{Payload} \\
        \midrule
        Relative Impact (\%) & 40.85\% & 32.72\% & 12.62\% & 7.80\% & 6.01\% \\
        \bottomrule
    \end{tabular}
\end{table}
\subsubsection{Model Choice Justification}
The polynomial regression model employed in this work is chosen to capture the complex aerodynamic interactions between drones operating within a shared skyway network. Prior empirical studies have demonstrated that polynomial models up to the third degree effectively capture aerodynamic behavior of drones operating in isolation while retaining interpretability~\cite{gupta2023aerodynamic, koo2018uav}. In contrast, we adopt a fourth-degree polynomial model to capture the higher-order nonlinearities that emerge from complex multi-drone interactions in a shared skyway network. In addition, the polynomial regression model provides clear insights into how specific factors influence inter-drone interference,
enabling super-providers (i.e., drone traffic management authorities) and stakeholders to easily interpret and justify operational decisions based on the model’s predictions~\cite{molnar2020interpretable}. For instance, our model identifies \emph{position} of a drone as a significant factor that influences delivery efficiency using a third-degree polynomial term. This type of insight is often difficult to extract from black-box models due to their limited interpretability~\cite{molnar2020interpretable}. Future work may explore complex models, but our results indicate that polynomial regression with a fourth degree captures complex aerodynamic interactions in a multi-drone skyway network and allows transparent decision-making in this context.

\subsection{Model Validation}
\label{sec:model_validation}

\begin{figure}[h!]
    \centering
    \begin{subfigure}{0.39\textwidth}
        \centering
        \includegraphics[width=\linewidth]{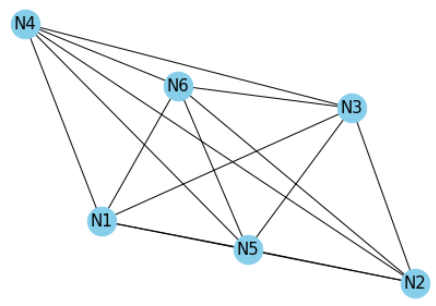}
        \caption{Dense network mimicking urban environment}
        \label{fig:first_dense_network}
    \end{subfigure}%
    \hspace{0.05\textwidth} 
    \begin{subfigure}{0.39\textwidth}
        \centering
        \includegraphics[width=\linewidth]{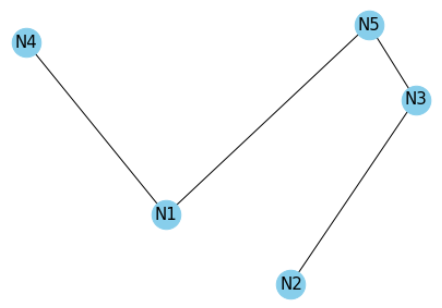}
        \caption{Sparse network mimicking suburban environment}
        \label{fig:second_sparse_network}
    \end{subfigure}
    \caption{Networks for validation trajectories.}
    \label{fig:validation_network}
\end{figure}
We validated the segment-level and node-level interference models using a dataset of 50 drone flights. These flights were conducted in two skyway network settings designed to test varying node densities and segment connectivity. The trajectories were excluded from the training dataset. The first network represents dense urban airspace with high node density (Fig.~\ref{fig:first_dense_network}). The second network mimics a suburban environment with fewer nodes and less connectivity (Fig.~\ref{fig:second_sparse_network}). In both settings, we directed 50 trajectories as solo and simultaneous flights. The segment-level model from Section~\ref{SegModel} predicts power consumption during simultaneous flights. Waiting times at nodes are estimated using the empirical model in Section~\ref{empirical model}. Delivery times are computed with Eq.~(\ref{DT_Interference}), accounting for node and segment-level interference. Model performance is evaluated using \emph{MAE}, \emph{MAPE}, \emph{MSE}, and \( R^2 \) using the following equations:

\begin{equation}
\begin{tabular}{c c c c}
\text{MAE} $= \frac{1}{n} \sum\limits_{i=1}^{n} \left| y_i - \hat{y}_i \right|$ & 
\text{MAPE} $= \frac{100}{n} \sum\limits_{i=1}^{n} \left| \frac{y_i - \hat{y}_i}{y_i} \right|$ &
\text{MSE} $= \frac{1}{n} \sum\limits_{i=1}^{n} \left( y_i - \hat{y}_i \right)^2$ &
$R^2 = 1 - \frac{\sum\limits_{i=1}^{n} \left( y_i - \hat{y}_i \right)^2}{\sum\limits_{i=1}^{n} \left( y_i - \bar{y} \right)^2}$
\end{tabular}
\label{eq:metrics}
\end{equation}

where \( \bar{y} \) denotes the mean of observed interference values \( y \).

\section{Experiments and Results}

We implement and evaluate the proposed methodology for analyzing inter-drone interference within a skyway network. Our assessment focuses on key performance metrics such as power consumption and delivery delays. We first conduct a series of solo drone flights of predefined trajectories and record their initial power consumption and delivery time. Then, we examine how simultaneous flights introduce interference effects in the form of increased power consumption and delivery delays. The experiments were designed to capture the impact of varying interference scenarios. We present a breakdown of the impact of these interference factors. We then evaluate the accuracy of our model.

\subsection{Dataset and Experiments Setup}
We collected a dataset of drone trajectories in an indoor testbed with 3D-printed models of Sydney's CBD \cite{Guo2023}. Each building was equipped with a wireless recharging and waiting pad as the primary landing and recharging infrastructure. Crazyflie 2.1\footnote{https://www.bitcraze.io/products/old-products/crazyflie-2-1/} drones with Qi charging and Lighthouse positioning decks were used. The drones traversed segments and recharged at intermediate nodes to complete delivery requests between source and destination rooftops. Two HTC Vive base stations enabled precise navigation, and two FANCO DC pedestal fans generated wind at different speeds and directions. Time-series data was collected using the CFLib Python Library. Table \ref{tab:data_parameters} lists the attributes recorded, including battery consumption and delivery time under inter-drone interference. The dataset consists of 1485 flights, including solo flights with independent drones and simultaneous flights in shared networks.

\begin{table}[htbp]
    \centering
    \rowcolors{1}{white}{gray!15}  
    \caption{Specifications of the drone model}
    \begin{tabularx}{0.45\textwidth}{X c} 
        \toprule
        \textbf{Specification}  & \textbf{Crazyflie 2.1} \\ 
        \midrule
        Weight (g)              & 27                    \\ 
        Max payload (g)         & 15                    \\ 
        Drone speed (m/s)       & 0.15                  \\ 
        Max flight time (min)   & 7                     \\ 
        Charging time (min)     & 40                    \\ 
        Takeoff size (mm) (WxHxD) & 92x92x29             \\ 
        \bottomrule
    \end{tabularx}
    \label{dronespecification}
\end{table}

\subsection{Analysis of Inter-Drone Interference Effects Across Scenario Variables}\label{Analysis}
In this section, we present the analysis of various factors that inform the impact of inter-drone interference on drones' delivery time and power consumption. We use the GUI to compare the impact of interferences when drones operate independently and simultaneously in the presence of peer drones. Table~\ref{dronespecification} outlines the specifications of the drone model used in our experiments. In what follows, we discuss the impact of various factors on the efficiency of drone delivery services.

Each experimental variable was studied in isolation, holding all other factors constant. Specifically, the dataset comprises balanced sample sizes for key scenario variables: drone configuration (Front-back: 495, Top-down: 495, Side-by-side: 495), separation (close: 495, moderate: 495, wide: 495), and wind conditions (none: 297, headwind low: 297, tailwind low: 297, headwind intense: 297, tailwind intense: 297). For the payload experiments, the dataset includes primarily three standardized payload conditions (none, light, heavy) with 429 flights each. The statistical analyses are derived from these consistent sample sizes, ensuring comparability and reliability across all presented results. The means and confidence intervals for each variable are summarized in Table~\ref{tab:compact_panel_power}.

\begin{table}[h!]
\centering
\caption{Means and 95\% confidence intervals for Power (\%) by scenario variable.}
\label{tab:compact_panel_power}
\begin{tabular}{lcc@{\hskip 18pt}lcc}
\toprule
\multicolumn{3}{c}{\textbf{Configuration}} & \multicolumn{3}{c}{\textbf{Separation}} \\
\cmidrule(lr){1-3} \cmidrule(lr){4-6}
\textbf{Level} & \textbf{Mean} & \textbf{95\% CI} & \textbf{Level} & \textbf{Mean} & \textbf{95\% CI} \\
front-back   & 16.98 & [15.50, 18.47] & close    & 17.33 & [15.82, 18.85] \\
side-by-side & 15.52 & [14.09, 16.94] & moderate & 17.74 & [16.18, 19.29] \\
top-down     & 20.26 & [18.59, 21.93] & wide     & 17.55 & [16.01, 19.09] \\
\addlinespace
\multicolumn{3}{c}{\textbf{Payload}}      & \multicolumn{3}{c}{\textbf{Wind}} \\
\cmidrule(lr){1-3} \cmidrule(lr){4-6}
\textbf{Level} & \textbf{Mean} & \textbf{95\% CI} & \textbf{Level} & \textbf{Mean} & \textbf{95\% CI} \\
heavy            & 20.60 & [18.85, 22.34] & headwind intense & 19.65 & [17.52, 21.77] \\
light            & 16.97 & [15.54, 18.40] & headwind low     & 16.72 & [14.82, 18.62] \\
none             & 14.67 & [13.39, 15.94] & none             & 15.75 & [13.69, 17.81] \\
                 &       &                & tailwind intense & 16.14 & [14.32, 17.96] \\
                 &       &                & tailwind low     & 19.04 & [17.09, 21.00] \\
\bottomrule
\end{tabular}
\end{table}

\subsubsection{Payload Effects}
\label{sec:payload_effects}
In drone delivery services, payload size varies significantly depending on the drone service provider’s requirements. This variation in payload introduces a critical factor that influences aerodynamic interference.  We examine how different payload configurations influence power consumption for all five drone flights. Heavier payloads are expected to amplify both segment-level and node-level interference effects leading to increased power consumption, and subsequently increased recharging time. We compare the magnitude increase between solo and simultaneous flight operations. 

\begin{figure}[htbp]
    \centering
    \begin{subfigure}[b]{0.47\textwidth}
        \centering
        \includegraphics[width=\textwidth]{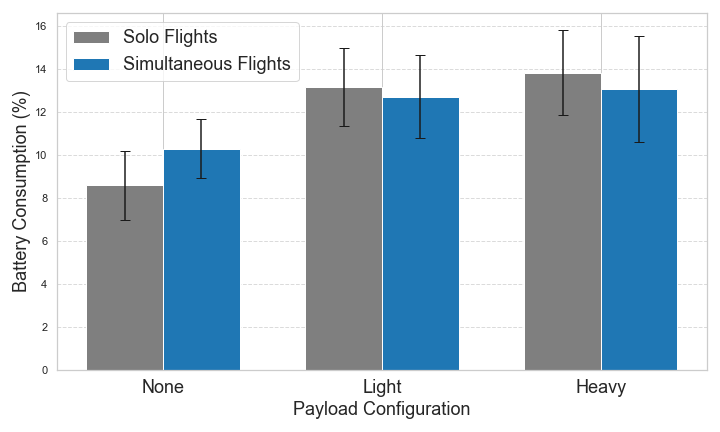}
        \caption{Drone 1}
        \label{fig:drone1_bar}
    \end{subfigure}
    \hfill
    \begin{subfigure}[b]{0.47\textwidth}
        \centering
        \includegraphics[width=\textwidth]{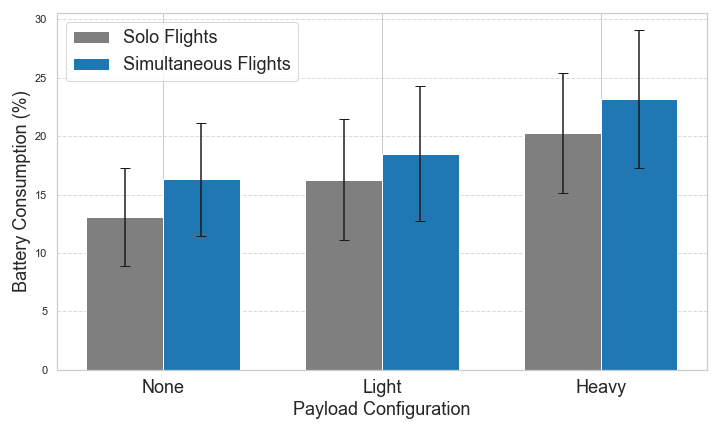}
        \caption{Drone 2}
        \label{fig:drone2_bar}
    \end{subfigure}
    
    \vspace{10pt}
    
    \begin{subfigure}[b]{0.47\textwidth}
        \centering
        \includegraphics[width=\textwidth]{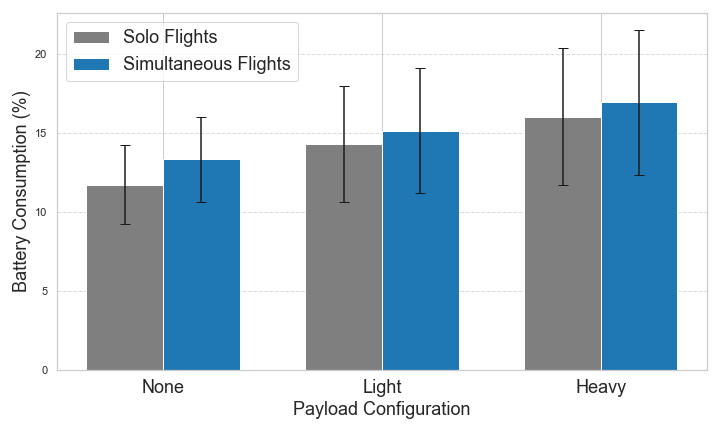}
        \caption{Drone 3}
        \label{fig:drone3_bar}
    \end{subfigure}
    \hfill
    \begin{subfigure}[b]{0.47\textwidth}
        \centering
        \includegraphics[width=\textwidth]{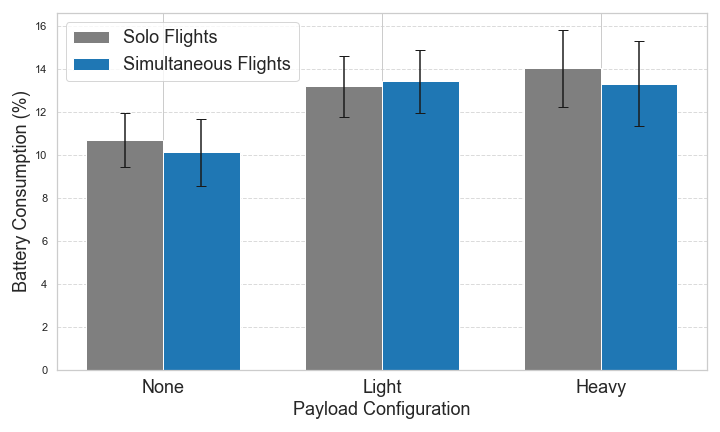}
        \caption{Drone 4}
        \label{fig:drone4_bar}
    \end{subfigure}
    
    \vspace{10pt}
    
    \begin{subfigure}[b]{0.47\textwidth}
        \centering
        \includegraphics[width=\textwidth]{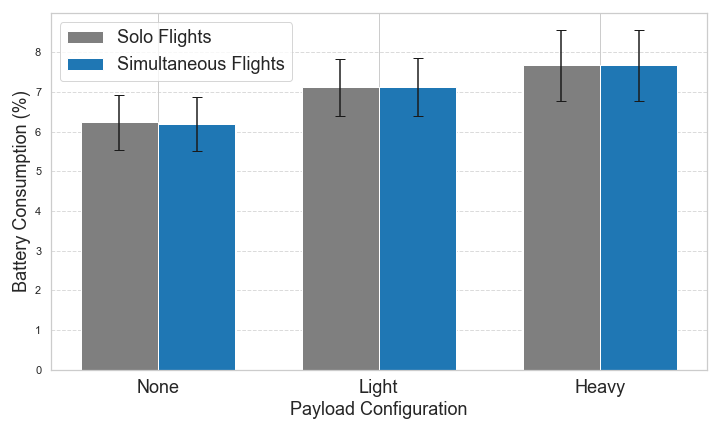}
        \caption{Drone 5}
        \label{fig:drone5_bar}
    \end{subfigure}

    \caption{Average battery consumption across payloads (none, light, heavy) for all drones, aggregated over all flight scenarios.}
    \label{fig:drones_payload}
\end{figure}

The five drones follow the expected trend of increased average power consumption with heavier payloads, with each weight category resulting in approximately 17\% higher consumption as shown in Fig. \ref{fig:drones_payload}. However, specific patterns emerge when comparing power consumption in solo versus simultaneous flights for each drone. Drones 2 and 3 experience a significant rise in power consumption due to interference. In contrast, Drone 5 shows no difference between solo and simultaneous flights as it does not encounter interference along its trajectory. Drones 1 and 4 exhibit higher power consumption during solo flights compared to simultaneous flights. This result is influenced by a tested \textit{front-and-back} formation, where Drone 1 leads Drone 2 and Drone 4 leads Drone 3. In these formations, the leading drones experience a \textit{tailwind} effect generated by the trailing drones, reducing power consumption during simultaneous flights.\looseness=-1

\begin{table}[htbp]
    \centering
    \rowcolors{1}{white}{gray!15}
    \caption{Comparison of Power Consumption, Recharging Time, and Delivery Delay for Different Payload Configurations for Drone 2 Flight Sample.}
    \label{table:payload_effects}
    \begin{tabular}{%
        >{\raggedright\arraybackslash}p{1.5cm}  
        >{\centering\arraybackslash}p{2.5cm}      
        >{\centering\arraybackslash}p{2.5cm}    
        >{\centering\arraybackslash}p{1.8cm}    
        >{\centering\arraybackslash}p{2.5cm}    
        >{\centering\arraybackslash}p{2cm}    
    }
        \toprule
        \textbf{Payload} & \textbf{Power Consumption (Solo) \%} & \textbf{Power Consumption (Simultaneous) \%} & \textbf{Recharging Time (Solo) (s)} & \textbf{Recharging Time (Simultaneous) (s)} & \textbf{Delivery Delay (s)} \\
        \midrule
        None     & 9.80  & 12.88 & 352 & 445 & 93  \\
        Light    & 10.81 & 14.23 & 383 & 486 & 103 \\
        Heavy    & 13.78 & 17.79 & 592 & 472 & 120 \\
        \bottomrule
    \end{tabular}
\end{table}

\begin{figure}[htbp]
    \centering
    \begin{minipage}[t]{0.48\textwidth}
        \centering
        \includegraphics[width=\textwidth]{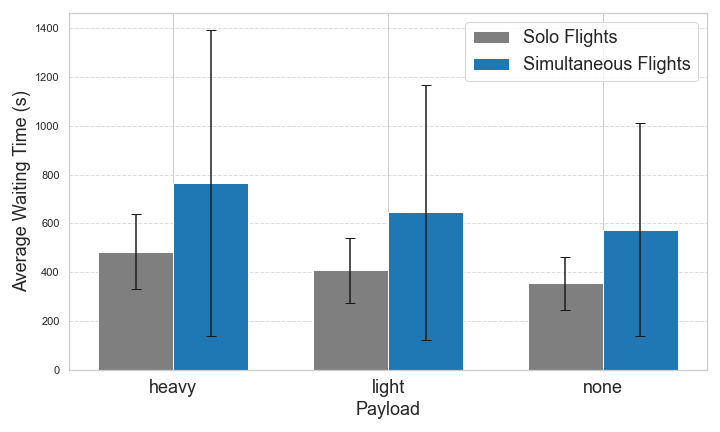}
        \caption{Average Waiting Time for Drones Based on Payload, Aggregated Across All Drones.}
        \label{fig:waiting_time_payload}
    \end{minipage}%
    \hfill
    \begin{minipage}[t]{0.48\textwidth}
        \centering
        \includegraphics[width=\textwidth]{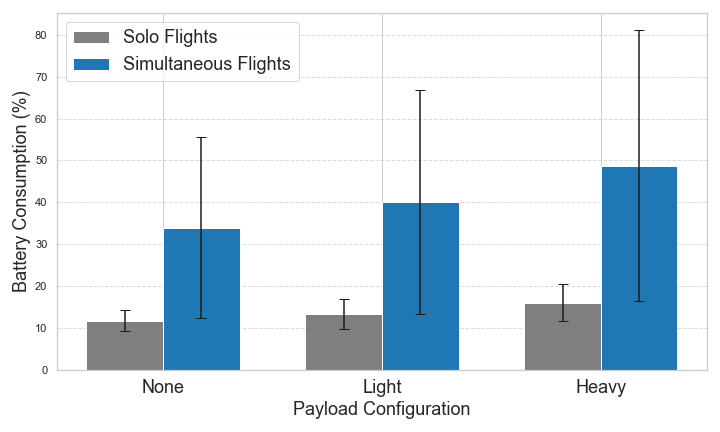}
        \caption{Power Consumption Due to Hovering for Different Payloads, Drone 3.}
        \label{fig:hovering_power_drone3}
    \end{minipage}
\end{figure}

Table~\ref{table:payload_effects} shows that segment-level interference increases power consumption by approximately 30\% to 35\% across all payloads in simultaneous flights. This rise is especially significant for Crazyflie drones due to their limited battery capacity. Higher power consumption extends recharging times by 26\% for lighter payloads and up to 34\% for heavier ones. The impact of heavier payloads further amplifies these interference effects. In addition to segment-level interference, payload weight directly influences node-level interference. Heavier payloads require more power to complete each flight, extending the recharging durations at nodes. As illustrated in Fig. \ref{fig:waiting_time_payload}, these extended recharging durations cascade into amplified waiting times at nodes. Heavier payloads intensify the resource contention at nodes as drones occupy the recharging pads for longer durations, beginning from 30\% up to 40\% when carrying the heaviest payload category.
This prolonged waiting time also has consequences for hovering drones. Prolonged hovering causes a notable increase in power consumption. In our set of drone flights Drone 3 has to hover on node N2 with all the waiting pads occupied by Drone 5 and 4. Fig. \ref{fig:hovering_power_drone3} demonstrates that higher payloads lead to significant power drain during these hovering payloads starting from 30\% at lighter payloads up to 40\% on the heavier payloads. Consequently, the drone requires additional time to recuperate this lost charge, which leads to further delays.

\begin{table}[htbp]
    \centering
    \rowcolors{1}{white}{gray!15}
    \caption{Delivery Delay for Drone 3 Across Payload Configurations.}
    \label{table:delivery_delay_drone3}
    \begin{tabular}{%
        >{\centering\arraybackslash}p{1.5cm}     
        >{\centering\arraybackslash}p{3.2cm}   
        >{\centering\arraybackslash}p{4.8cm}     
        >{\centering\arraybackslash}p{3.2cm}   
    }
        \toprule
        \textbf{Payload} & \textbf{Hovering Time (s)} & \textbf{Hovering Power Consumption (\%)} & \textbf{Recharging Time (s)} \\
        \midrule
        None  & 352 & 32.7 & 964  \\
        Light & 383 & 45.2 & 1264 \\
        Heavy & 472 & 55.3 & 1506 \\
        \bottomrule
    \end{tabular}
    \label{table:delivery_delay_drone3}
\end{table}

Table~\ref{table:delivery_delay_drone3} illustrates the compounded effect of payload on recharging and hovering where each contributes to additional delay. Heavier payloads incur prolonged recharging and require more hovering, increasing the total delivery delay. These delays impact service providers' ability to meet time-sensitive demands. Our results emphasize on the importance of managing payload distributions across service providers' operational plans. Effective planning allows service providers to alleviate a notable amount of delays and operational costs.

\subsubsection{Drone Formation, Position, and Separation Effects}
\label{sec:drone_formation}
In multi-drone delivery services, different formations are often required based on the flight paths and spatial constraints of the skyway network. Drones may need to fly in side-by-side, top-down, or front-back formations, depending on factors such as airspace availability, delivery priority, or recharging schedules. These formations, and the positions drones occupy within them introduce distinct aerodynamic interactions. We examine the effects of different drone formations and positions within those formations on power consumption. Each formation introduces distinct aerodynamic interactions, which may influence the interference effects observed during simultaneous flight operations. These effects inform the eventual delivery performance of drones with regard to the formation employed.

\paragraph{Side-by-Side Formation}

\begin{figure}[H]
    \centering
    \begin{minipage}[htbp]{0.48\textwidth}
        \centering
        \includegraphics[width=\textwidth]{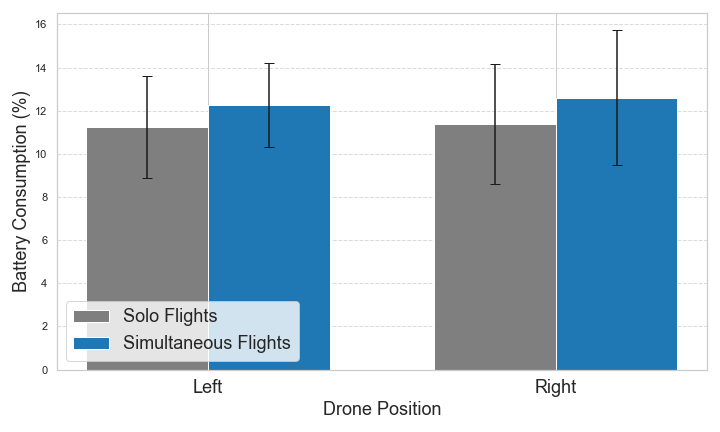}
        \caption{Average Battery Consumption by Position for Solo vs Simultaneous Flights in Side-by-Side Formation.}
        \label{fig:avg_battery_by_position}
    \end{minipage}%
    \hfill
    \begin{minipage}[htbp]{0.48\textwidth}
        \centering
        \includegraphics[width=\textwidth]{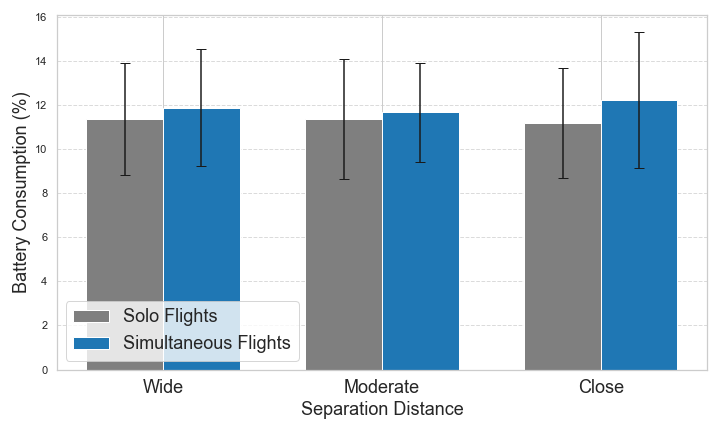}
        \caption{Average Battery Consumption by Separation Distance for Solo vs Simultaneous Flights in Side-by-Side Formation.}
        \label{fig:avg_battery_by_separation}
    \end{minipage}
\end{figure}

As shown in Fig.~\ref{fig:avg_battery_by_position}, on average both the left and right positions exhibit increased power consumption compared to solo flights. There is no significant difference observed between the two positions. Furthermore, as depicted in Fig.~\ref{fig:avg_battery_by_separation}, decreasing the separation distance between the drones amplifies this interference, leading to higher power consumption.

\paragraph{Front-Back Formation}

\begin{figure}[H]
    \centering
    \begin{minipage}[htbp]{0.48\textwidth}
        \centering
        \includegraphics[width=\textwidth]{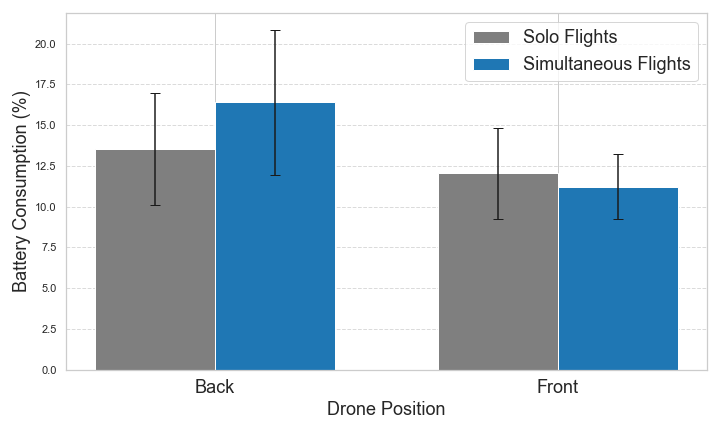}
        \caption{Average Battery Consumption by Position for Solo vs Simultaneous Flights in Front-Back Formation.}
        \label{fig:avg_battery_by_position_front_back}
    \end{minipage}%
    \hfill
    \begin{minipage}[htbp]{0.48\textwidth}
        \centering
        \includegraphics[width=\textwidth]{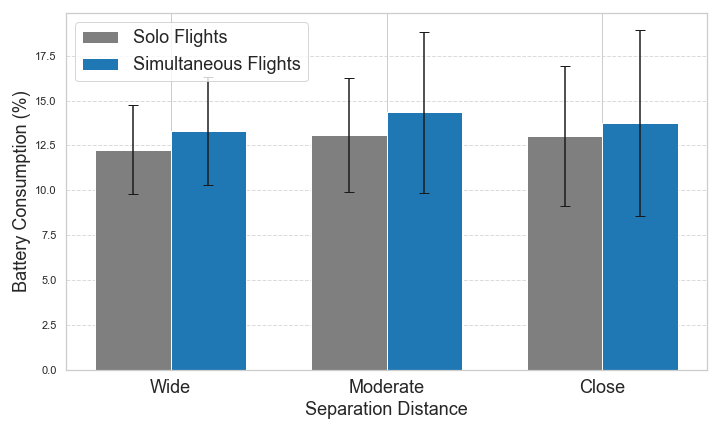}
        \caption{Average Battery Consumption by Separation Distance for Solo vs Simultaneous Flights in Front-Back Formation.}
        \label{fig:avg_battery_by_separation_front_back}
    \end{minipage}
\end{figure}

The results averaged across all flight scenarios reveal distinct differences in power consumption based on drone position (see Fig.~\ref{fig:avg_battery_by_position_front_back}). In simultaneous flights, the leading drone consistently uses less power. This reduction occurs because the trailing drone generates a tailwind that assists the lead drone, reducing its energy demands. In contrast, the trailing drone experiences higher power consumption. This increase results from aerodynamic drag as it operates in the wake of the lead drone, forcing it to expend more energy to maintain stability and speed. Figure~\ref{fig:avg_battery_by_separation_front_back} further demonstrates that closer separation distances between drones amplify these effects. It is important to note that this impact is small due to the compact scale of our drones. Nevertheless, the trend suggests that strategizing separation could increase the efficiency advantage for the leading drone.

\paragraph{Top-Down Formation}

\begin{figure}[H]
    \centering
    \begin{minipage}[htbp]{0.48\textwidth}
        \centering
        \includegraphics[width=\textwidth]{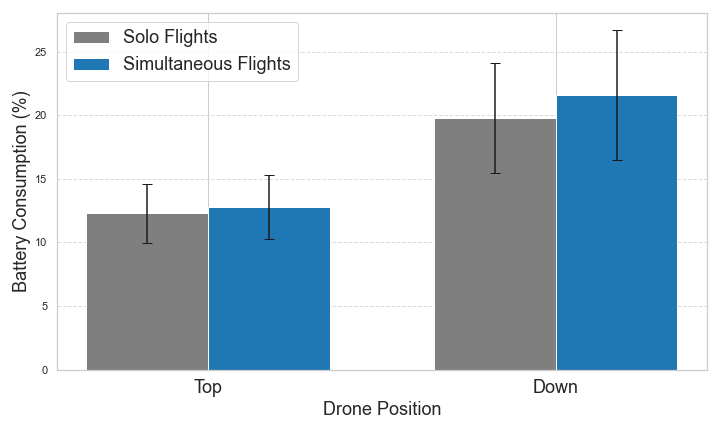}
        \caption{Average Battery Consumption by Position for Solo vs Simultaneous Flights in Top-Down Formation.}
        \label{fig:avg_battery_by_position_top_down}
    \end{minipage}%
    \hfill
    \begin{minipage}[htbp]{0.48\textwidth}
        \centering
        \includegraphics[width=\textwidth]{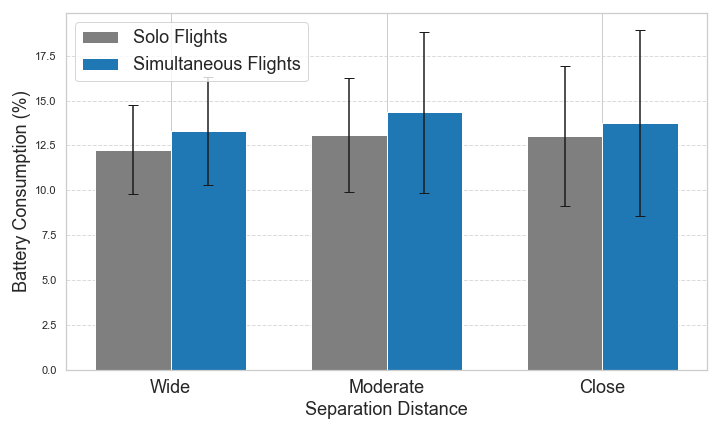}
        \caption{Average Battery Consumption by Separation Distance for Solo vs Simultaneous Flights in Top-Down Formation.}
        \label{fig:avg_battery_by_separation_top_down}
    \end{minipage}
\end{figure}

This formation results in increased power consumption for the bottom drone due to the downwash generated by the upper drone. As illustrated in Fig.~\ref{fig:avg_battery_by_position_top_down}, the bottom drone consistently consumes more power during simultaneous flights compared to solo flights. In contrast, the top drone experiences less disturbance. Furthermore, as demonstrated in Fig.~\ref{fig:avg_battery_by_separation_top_down}, increasing the separation distance between the two drones reduces the aerodynamic influence, lowering the bottom drone's power consumption.

\begin{table}[htbp]
    \centering
    \rowcolors{2}{white}{gray!15}
    \caption{Power Consumption, Recharging Time, and Delivery Delay by Drone Formation and Position.}
    \label{tab:delivery_times_table_formation}
    \begin{tabular}{%
        >{\raggedright\arraybackslash}p{1.8cm}  
        >{\raggedright\arraybackslash}p{1.2cm}    
        >{\centering\arraybackslash}p{2cm}      
        >{\centering\arraybackslash}p{2.3cm}    
        >{\centering\arraybackslash}p{1.5cm}    
        >{\centering\arraybackslash}p{2cm}    
        >{\centering\arraybackslash}p{1.8cm}    
    }
        \toprule
        \textbf{Formation Type} & \textbf{Drone Position} & \textbf{Power Consumption (Solo) \%} & \textbf{Power Consumption (Simultaneous) \%} & \textbf{Recharging Time (Solo) (s)} & \textbf{Recharging Time (Simultaneous) (s)} & \textbf{Delivery Time Net Difference (s)} \\
        \midrule
        Side-by-Side  & Left    & 11.66 & 11.99 & 409 & 418 & 9 \\ 
                      & Right   & 10.43 & 11.16 & 364 & 372 & 8 \\ 
        Front-Back    & Front   & 12.40 & 10.04 & 431 & 360 & -71 \\ 
                      & Back    & 10.71 & 11.78 & 380 & 412 & 32 \\ 
        Top-Down      & Top     & 18.22 & 18.36 & 605 & 610 & 5 \\ 
                      & Bottom  & 12.25 & 13.04 & 426 & 450 & 24 \\ 
        \bottomrule
    \end{tabular}
    \label{tab:delivery_times_table_formation}
\end{table}

The three different formations evaluated have varying implications of delivery timeliness as shown in Table \ref{tab:delivery_times_table_formation}. In the front-back configuration, the lead drone consumes 19\% less power than in solo flights, resulting in a 16\% reduction in recharging time and a 71-second decrease in delivery time. Meanwhile, the trailing drone experiences a 10\% increase in power consumption, extending its delivery time by 32 seconds. This formation can be advantageous for time-sensitive deliveries. In the side-by-side formation, power consumption and delivery time show minimal differences between the left and right positions, with only slight increases in both metrics. This consistency makes the side-by-side formation well-suited for scenarios requiring uniform efficiency across drones. In the top-down configuration, aerodynamic downwash from the top-positioned drone raises the bottom drone’s power consumption by 6\%, causing a 24-second delivery delay, making this formation the least ideal in our results.

These findings emphasize the strategic value of formation selection based on airspace and operational needs. The front-back formation benefits urgent deliveries, while the side-by-side formation provides balanced, consistent performance. Selecting formations that match specific goals can help reduce delays for service providers.

\subsubsection{Wind Condition Effects}

External environmental conditions, particularly wind, play a significant role in the efficiency of drone operations. Wind conditions such as headwinds and tailwinds have a notable impact on power consumption. Headwinds generally increase power consumption, while tailwinds may provide aerodynamic advantages by reducing energy requirements. In the context of inter-drone interference, wind conditions can either exacerbate or alleviate interference effects, particularly when drones fly in close proximity.

\begin{figure}[htbp]
    \centering
    \begin{minipage}[t]{0.49\textwidth}
        \centering
        \includegraphics[width=\textwidth]{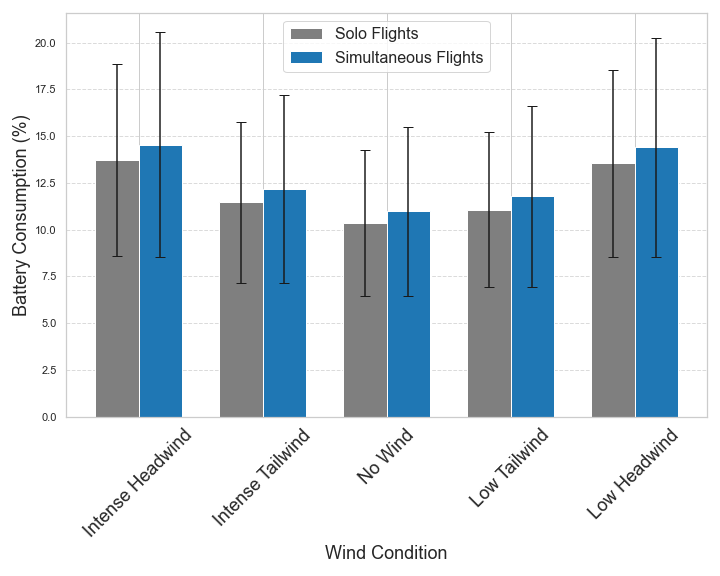}
        \caption{Average battery consumption under different wind conditions, aggregated across all drones and scenario variables.}
        \label{fig:bar_graph_wind}
    \end{minipage}
    \hfill
    \begin{minipage}[t]{0.49\textwidth}
        \centering
        \includegraphics[width=\textwidth]{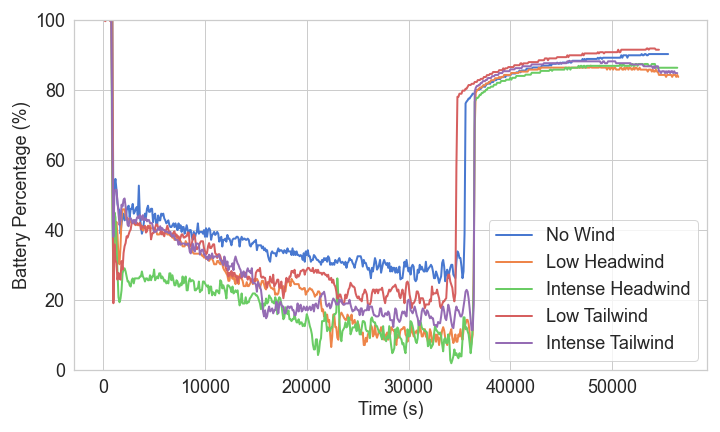}
        \caption{Time-series comparison of battery consumption under different wind conditions.}
        \label{fig:time_series_wind}
    \end{minipage}
\end{figure}

As shown in Fig.~\ref{fig:bar_graph_wind}, headwinds consistently lead to higher power consumption, whereas tailwinds result in reduced energy usage. The intensity of the wind amplifies these effects, with stronger winds having a more pronounced impact. Both solo and simultaneous flights are influenced by wind conditions, though simultaneous flights show slightly higher consumption due to the combined aerodynamic interference and wind effects. In addition, we analyze a sample of flights under five wind conditions: no wind, low headwind, intense headwind, low tailwind, and intense tailwind. Figure~\ref{fig:time_series_wind} provides a time-series comparison of battery consumption over these conditions. The presence of wind results in erratic power consumption patterns due to the drone’s need to compensate for unstable flight. Under tailwind conditions, battery levels decrease more gradually, as the wind assists the drone's forward movement, reducing strain on the motors.
\begin{table}[t!]
    \centering
    \rowcolors{2}{white}{gray!15}
    \caption{Power Consumption, Recharging Time, and Delivery Delay by Wind Condition for Simultaneous Flights.}
    \label{tab:delivery_times_table_wind}
    \begin{tabular}{%
        >{\raggedright\arraybackslash}p{3cm}   
        >{\centering\arraybackslash}p{3.5cm}     
        >{\centering\arraybackslash}p{3cm}     
        >{\centering\arraybackslash}p{3.7cm}       
    }
        \toprule
        \textbf{Wind Condition} & \textbf{Power Consumption (\%)} & \textbf{Recharging Time (s)} & \textbf{Delivery Time Increase (s)} \\
        \midrule
        No Wind           & 8.81  & 316  & 0   \\ 
        Low Headwind      & 9.75  & 350  & 34  \\ 
        Intense Headwind  & 11.67 & 409  & 93  \\ 
        Low Tailwind      & 9.31  & 334  & 18  \\ 
        Intense Tailwind  & 9.42  & 338  & 22  \\ 
        \bottomrule
    \end{tabular}
    \label{tab:delivery_times_table_wind}
\end{table}
These results confirm that wind consistently increases power consumption, regardless of its direction or speed (see Table~\ref{tab:delivery_times_table_wind}). Headwinds have the greatest impact on power consumption. Intense headwinds cause a 32\% rise in power consumption, extend recharging time by nearly 30\%, and delay delivery by 93 seconds. Tailwinds also increase power consumption, though less severely. Intense tailwinds lead to a 7\% increase in power consumption and a 22-second delivery delay. This difference suggests a strategic opportunity. Service providers should integrate wind forecasts into operational planning to reduce disruptions. Aligning flight paths with tailwinds, as shown by our results, can help offset negative impacts on power consumption.

\subsection{Model Validation Results}
The following results demonstrate the performance of the proposed interference model on the validation dataset. The model was evaluated across various interference scenarios to gauge the accuracy in predicting power consumption and delivery time. We evaluate the model accuracy using the evaluation metrics (i.e., MAPE, MAE, MSE, and $R^2$) discussed in Section~\ref{sec:model_validation}. 
\begin{table}[h!]
    \centering
    \caption{Performance of the Proposed Interference Model on the Validation Set.}
    \label{tab:model_performance}
    \begin{tabularx}{\textwidth}{p{3cm}p{1cm}XXXX}
        \toprule
        & & \multicolumn{4}{c}{\textbf{Testing}} \\
        \cmidrule(lr){3-6}
        \textbf{Model} & \textbf{Metric} & \textbf{MAE} & \textbf{MSE} & \textbf{MAPE} & \textbf{\hspace{0.1cm}$R^2$} \\
        \midrule
        \multirow{2}{*}{\makecell{\textbf{Proposed}\\\textbf{Interference Model}}} 
            & Power & 1.62 & 4.25 & 4.63\% & 0.98 \\
            & Time & 198.86 & 91370.06 & 10.79\% & 0.76 \\
        \bottomrule
    \end{tabularx}
\end{table}
 
\par
As shown in Table~\ref{tab:model_performance}, the model achieves MAE, MSE, and MAPE scores of 1.62, 4.25, and 4.63, respectively, for predicting power consumption due to interference. These low error scores demonstrate the model’s ability to make accurate predictions with a minimal error margin relative to observed values. The high $R^2$ value of 0.99 further supports this, indicating that the model effectively captures nearly all variance in power consumption data, showing strong predictive reliability. For delivery time estimation impacted by inter-drone interference, the model yields a MAE and MSE of 198.86 and 91370.06, respectively. These higher values reflect the complexity of accurately predicting delivery times affected by cumulative recharging needs. However, the $R^2$ value of 0.76 indicates that, despite these complexities, the model captures a substantial portion of the variance in delivery times, particularly representing the broader trends in the data.

\begin{figure}[h!]
    \centering
    \includegraphics[width=1.0\linewidth]{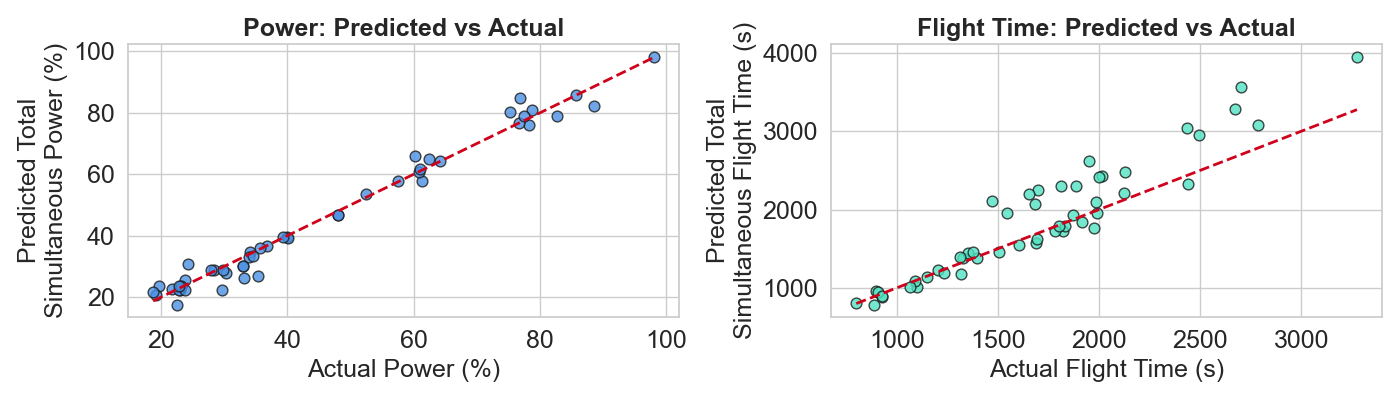}
    \caption{Model Predictions vs. Actual Measurements of Power Consumption and Flight Times for Validation Trajectories.}
    \label{fig:trajectory_overlay}
\end{figure}

In Fig.~\ref{fig:trajectory_overlay}, the 45-degree reference line indicates ideal accuracy, where predicted values match actual values precisely. For power consumption, data points closely align with this line, demonstrating the model’s accuracy and low prediction bias. The narrow spread around the line reflects consistent performance across scenarios. 85\% of power predictions fall within a ±5\% interval of the actual values. In contrast, flight time predictions exhibit a modest spread. Deviations appear mainly in high node-level interference scenarios, particularly instances involving prolonged hovering. In more severe interference scenarios, higher power demands affect the accuracy of recharging time estimates. 
\begin{figure}[h!]
    \centering
    \includegraphics[width=1.0\linewidth]{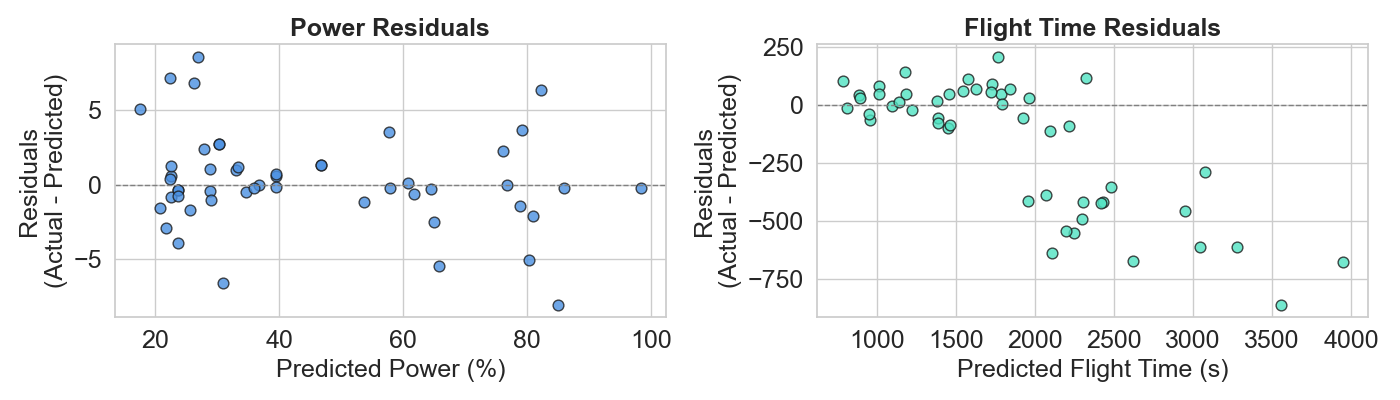}
    \caption{Scatter Plot of Prediction Residuals for Power Consumption and Flight Time.}
    \label{fig:residual_scatter}
\end{figure}

The prediction residuals points are distributed evenly around the zero line in Fig.~\ref{fig:residual_scatter}. This distribution indicates that the model does not consistently overestimate or underestimate power across the prediction range. The absence of a clear pattern suggests high accuracy in power prediction with minimal directional bias. In contrast, the flight time residuals show larger deviations at higher predicted values. This trend occurs mainly in high node-level interference scenarios, especially those with extended hovering times. In these cases, the model underestimates recharging time due to the increased power demands. This underestimation means that actual recharging times are longer than the model’s predictions.\looseness=-1

In summary, while the proposed model demonstrates strong predictive performance for power consumption, delivery time estimation initially faces higher error rates under conditions of increased node-level interference. Prediction errors are most pronounced early in missions due to uncertain queueing behaviors at recharging nodes. However, as drones approach their destinations, node-level interference reduces significantly. Straightforward segment traversal becomes more dominant in these later stages. Referring back to our validation set results, our model predicts these segments with high accuracy (Table~\ref{tab:comparison_results} ($R^2$ = 0.98)). As a result, prediction accuracy improves as the mission progresses. Consequently, delivery time predictions improve notably as the mission progresses. Addressing these initial uncertainties through more advanced queueing and simulation-based models is identified as a key avenue for future research.

\subsection{Performance Evaluation}
We conduct our experimental evaluation in two parts: (a) We ran a set of experiments to evaluate the performance of our proposed model against four state-of-the-art approaches from the literature.
(b) We conducted ablation experiments to quantitatively assess how each feature contributes to our proposed model’s predictive performance.

\begin{table}[t!]
    \centering
    \caption{Hyperparameters for Comparative Models.}
    \label{tab:model_hyperparameters}
    \begin{tabular}{ll|ll}
        \toprule
        \textbf{Elastic Net} & & \textbf{XGBoost} & \\
        $\alpha \in [0, 1]$ & & No. Trees $[100, 1000]$ & \\
        $L1$ Ratio $\in \{10^{-3}, 10^{-2}, 10^{-1}\}$ & & Max Depth $[2, 7]$ & \\
        & & Learning Rate $\{0.1, 0.3, 0.5\}$ & \\
        & & ColSampleByTree $\{0.5, 0.8, 1\}$ & \\
        \midrule
        \textbf{Random Forest} & & \textbf{SVR} & \\
        No. Trees $[100, 1000]$ & & $C \in \{0.1, 1, 10\}$ & \\
        Max Depth $[2, 7]$ & & Kernel $\in \{$rbf, linear$\}$ & \\
        & & $\epsilon \in \{0.01, 0.1\}$ & \\
        \bottomrule
    \end{tabular}
\end{table}
\subsubsection{Performance Effectiveness}
\label{sec:performance_effectiveness}
We evaluate the performance of our proposed interference model against state-of-the-art machine learning approaches: Elastic Net, XGBoost, Random Forest Regression, and Support Vector Regression (SVR). 
All models are implemented using hyperparameters identified as optimal for predicting delivery delays in previous studies~\cite{sarkar-2024,prasetia2022,dai2024}. Table~\ref{tab:model_hyperparameters} summarizes the hyperparameter configurations used for each model. To ensure a fair comparison, all models are evaluated under identical experimental conditions, where training and testing are conducted on the same dataset (Section~\ref{sec:model_validation}). We report the effectiveness of our model in terms of MAE, MSE, MAPE, $R^2$, and prediction time.
\begin{table}[htbp]
    \centering
    \caption{Comparison of Model Performance on the Validation Set}
    \label{tab:comparison_results}
    \begin{tabularx}{\textwidth}{p{3cm}p{1.3cm}XXXXc}
        \toprule
        \textbf{Model} & \textbf{Metric} & \textbf{MAE} & \textbf{MSE} & \textbf{MAPE} & \textbf{\hspace{0.1cm}$R^2$} & \textbf{Prediction Time (ms)} \\
        \midrule
        \multirow{2}{*}{\makecell{\textbf{Proposed}\\\textbf{Interference Model}}} 
            & Power & 1.62 & 4.25 & 4.63\% & 0.993
            & 0.32 \\
            & Time & \textbf{199.00} & \textbf{91370.00} & \textbf{10.79\%} & \textbf{0.760} & \\
        \midrule
        \multirow{2}{*}{\textbf{XGBoost}} 
            & Power & \textbf{0.29} & \textbf{0.43} & \textbf{0.88\%} & \textbf{0.999} & 1.6 \\
            & Time & 516.72 & 489577.86 & 27.04\% & -0.289 & \\
        \midrule
        \multirow{2}{*}{\textbf{Elastic Net}} 
            & Power & 1.47 & 3.38 & 4.22\% & 0.993 & 0.22 \\
            & Time & 527.75 & 494407.17 & 28.01\% & -0.301 & \\
        \midrule
        \multirow{2}{*}{\textbf{Random Forest}} 
            & Power & 1.02 & 2.21 & 3.01\% & 0.995 & 91.5 \\
            & Time & 521.53 & 488578.13 & 27.58\% & -0.286 & \\
        \midrule
        \multirow{2}{*}{\makecell{\textbf{Support Vector}\\\textbf{Regression}}} 
            & Power & 1.17 & 2.01 & 3.43\% & 0.996 & 13.3 \\
            & Time & 520.85 & 487408.20 & 27.53\% & -0.283 & \\
        \bottomrule
    \end{tabularx}
\end{table}

As shown in Table~\ref{tab:comparison_results}, our proposed model achieves comparable performance in comparison to other approaches for predicting power consumption. While performance is similar, note that the proposed model offers interpretability in comparison to other machine learning models. Polynomial regression model expresses the effect of each variable through direct coefficients, allowing assessment of factors such as payload weight or flight formation on the impact of interference~\cite{molnar2020interpretable}. In contrast, the other approaches formulate predictions based on ensembles, regularization, or kernel transformations that obscure the individual impact of variables. While these methods offer marginal improvements in accuracy, the proposed model achieves a better balance between predictive performance and interpretability that places it ahead in terms of overall usability. In addition, the proposed model reduces prediction time by approximately $\approx$98\% on average compared to other baseline methods. High predictive speed is especially crucial in large-scale deployments, where thousands of drone flights must be monitored to avoid cascading delays and service disruptions. Furthermore, the proposed model outperforms other baseline models in predicting delivery time. As shown in Table~\ref{tab:comparison_results}, it achieves the lowest MAE (199.00) and MSE (91370.00), along with the highest $R^2$ score (0.760), indicating a significantly better fit for delivery time prediction. This is because the proposed model captures the impact of interferences occurring at the node level due to forced hovering or waiting of drones at the node. The proposed model addresses these effects using the empirical model proposed in Section~\ref{sec:node_level_interference}. These findings demonstrate the effectiveness of our proposed model in comparison to other state-of-the-art models.

\subsubsection{Ablation Study}
\begin{table}[t!]
    \centering
    \caption{Results from Ablation Experiments}
    \label{tab:ablation_results}
    \begin{tabularx}{\textwidth}{p{2.5cm}l *{4}{>{\centering\arraybackslash}X} *{4}{>{\centering\arraybackslash}X}}
        \toprule
        \multirow{2}{*}{\textbf{Excluded Variable}} & & \multicolumn{4}{c}{\textbf{Power}} & \multicolumn{4}{c}{\textbf{Flight Time}} \\
        \cmidrule(lr){3-6} \cmidrule(lr){7-10}
         & & \textbf{MAE} & \textbf{MSE} & \textbf{MAPE} & $\mathbf{R^2}$ & \textbf{MAE} & \textbf{MSE} & \textbf{MAPE} & $\mathbf{R^2}$ \\
        \midrule
        \end{tabularx}
    \rowcolors{2}{white}{gray!15}
    \begin{tabularx}{\textwidth}{p{2.5cm}l *{4}{>{\centering\arraybackslash}X} *{4}{>{\centering\arraybackslash}X}}
        None (all features) & & 1.53 & 3.40 & 4.35\% & 0.99 & 196.94 & 90047.81 & 10.57\% & 0.76 \\
        formation           & & 1.53 & 3.40 & 4.35\% & 0.99 & 196.94 & 90047.81 & 10.57\% & 0.76 \\
        position            & & 2.28 & 8.82 & 6.45\% & 0.98 & 213.44 & 95966.74 & 12.20\% & 0.75 \\
        separation          & & 1.54 & 3.43 & 4.36\% & 0.99 & 197.01 & 90155.21 & 10.58\% & 0.76 \\
        payload             & & 1.35 & 3.84 & 4.04\% & 0.99 & 201.07 & 102912.60 & 10.72\% & 0.73 \\
        wind                & & 1.53 & 3.39 & 4.34\% & 0.99 & 196.86 & 89964.88 & 10.57\% & 0.76 \\
        \bottomrule
    \end{tabularx}
\end{table}

We conducted an ablation study to assess the sensitivity of our model against each feature variable. In this respect, we retrained the model by removing one feature at a time and assessed the performance drop using the same evaluation metrics as used in the model with all features. We then validate these effects against the trends observed in our experimental analysis presented in Section~\ref{Analysis}. 

The results from ablation experiments are reported in Table~\ref{tab:ablation_results}. Removing \emph{position} from the feature set results in a significant increase in prediction errors across all metrics (power MAE increases from 1.53 to 2.28, and flight time MAE increases from 196.94 to 213.44). These findings align with the empirical analysis presented in Section~\ref{sec:drone_formation}. Our analysis revealed that the position of the drone relative to the peer drone in close proximity directly impacts power consumption and flight time. Furthermore, as shown in Table~\ref{tab:ablation_results}, excluding \emph{payload} increases flight time MAE from 196.94 to 201.07. This result reflects the experimental observation in Section~\ref{sec:payload_effects}, where payloads directly impact the effect of interferences and consequently delivery time. For instance, a drone with a heavier payload consumes increased battery power due to the effect of turbulence from a nearby drone compared to a drone with a lighter payload. This leads to longer recharging durations and increased delivery time. In contrast, the limited effect observed when excluding formation (no change in MAE), separation (power MAE increases from 1.53 to only 1.54), or wind (no notable changes) corresponds closely with empirical evidence suggesting their effects are either modest or implicitly captured through position and payload. The consistency between the ablation results and experimental analysis in Section~\ref{Analysis} supports the validity of our selected features and generalizability of the proposed model.

\section{Conclusion}
We propose a novel model that informs the impact of inter-drone interference on the Quality of Service (QoS) of drone services in multi-drone skyway networks. We collected a dataset of multiple drone trajectories in an indoor drone testbed. We then conducted an in-depth analysis of inter-drone interferences as drones traverse the airspace to their destination while recharging at intermediate nodes. We then developed a model that informs the impact of interference on power consumption and delivery time when drones operate simultaneously with peer drones. The results show that the relative position of a drone to peer drones significantly affects the interference impact on drone services' QoS. We then evaluate our proposed model for predicting interference effects compared to the state-of-the-art approaches. Our results indicate the proposed model accurately predicts the impact of interferences on drone efficiency while ensuring low computational cost. 
It is important to note that factors such as payload volume, drone speed, and shape are outside the scope of our current model. Our future work will involve incorporating these additional factors to improve the model’s accuracy and applicability across diverse scenarios. Additionally, we aim to leverage these insights to develop inter-drone interference resolution strategies for the efficient operation of drones in multi-drone networks.

\section*{Acknowledgements}

This research was partly made possible by LE220100078 and DP220101823 grants from the Australian Research Council. The statements made herein are solely the responsibility of the authors.

\bibliographystyle{ACM-Reference-Format}
\bibliography{main}

\end{document}